\documentclass[12pt]{article}
\usepackage{amsfonts}
\usepackage{amssymb}
\usepackage{graphicx}
\usepackage{amsmath}
\usepackage{makeidx}
\usepackage{indentfirst}
\usepackage[T1]{fontenc}
\usepackage[utf8]{inputenc}
\usepackage{authblk}

\newcounter{resultnum}[section]
\newcounter{conclusionnum}[section]
\newcounter{conditionnum}[section]
\newcounter{conjecturenum}[section]
\newcounter{examplenum}[section]
\newcounter{exercisenum}[section]
\newcounter{lemmanum}[section]
\newcounter{notationnum}[section]
\newcounter{theoremnum}[section]
\newcounter{definitionnum}[section]
\newcounter{corollarynum}[section]
\newcounter{remarknum}[section]
\newcounter{propositionnum}[section]
\newcounter{acknowledgementnum}[section]
\newcounter{algorithmnum}[section]
\newcounter{axiomnum}[section]
\newcounter{casenum}[section]
\newcounter{claimnum}[section]
\newcounter{summarynum}[section]
\newcounter{problemnum}[section]
\begin{document}

\title{Starobinsky Inflation and Dark Energy and Dark Matter Effects from
Quasicrystal Like Spacetime Structures}
\date{May 7, 2018}
\author{${}$ \\
Raymond Aschheim\\
{\small \textit{Quantum Gravity Research; 101 S. Topanga Canyon Blvd \#
1159. Topanga, CA 90290, USA}} \\
{\small \textit{email: raymond@quantumgravityresearch.org }}\\
${}$ \\
Lauren\c{t}iu Bubuianu\\
{\small \textit{TVR Ia\c si, 33 L. Catargi street; and University Apollonia,
2 Muzicii street, 700107 Ia\c si, Romania}} \\
{\small \textit{email: laurentiu.bubuianu@tvr.ro }}\\
${}$ \\
Fang Fang\\
{\small \textit{Quantum Gravity Research; 101 S. Topanga Canyon Blvd \#
1159. Topanga, CA 90290, USA}} \\
{\small \textit{email: Fang@quantumgravityresearch.org }}\\
${}$ \\
Klee Irwin \\
{\small \textit{Quantum Gravity Research; 101 S. Topanga Canyon Blvd \#
1159. Topanga, CA 90290, USA}} \\
{\small \textit{email: klee@quantumgravityresearch.org }}\\
${}$ \\
Vyacheslav Ruchin \\
{\small \textit{Heinrich-Wieland-Str. 182,\ 81735 M\"{u}nchen, Germany }} \\
{\small \textit{email: v.ruchin-software@freenet.de }}\\
${}$ \\
Sergiu I. Vacaru\thanks{
\textit{for postal correspondence with this author use:\ } 67 Lloyd Street
South, Manchester, M14 7LF, the UK} \\
{\small \textit{\ Physics Department, California State University Fresno,
Fresno, CA 93740, USA}} \\
{\small and \textit{University "Al. I. Cuza" Ia\c si, Project IDEI, Ia\c si,
Romania }} \\
{\small \textit{email: sergiu.vacaru@gmail.com and
sergiuvacaru@mail.csufresno.edu}} }
\maketitle

\begin{abstract}
The goal of this work on mathematical cosmology and geometric methods in
modified gravity theories, MGTs, is to investigate Starobinsky-like
inflation scenarios determined by gravitational and scalar field
configurations mimicking quasicrystal, QC, like structures. Such spacetime
aperiodic QCs are different from those discovered and studied in solid state
physics but described by similar geometric methods. We prove that a
inhomogeneous and locally anisotropic gravitational and matter field
effective QC mixed continuous and discrete "ether" can be modelled by exact
cosmological solutions in MGTs and Einstein gravity. The coefficients of
corresponding generic off-diagonal metrics and generalized connections
depend (in general) on all spacetime coordinates via generating and
integration functions and, certain smooth and discrete parameters. Imposing
additional nonholonomic constraints, prescribing symmetries for generating
functions and solving the boundary conditions for integration functions and
constants, we can model various nontrivial torsion QC structures or extract
cosmological Levi-Civita configurations with diagonal metrics reproducing de
Sitter (inflationary) like and other type homogeneous inflation and
acceleration phases. Finally, we speculate how various dark energy and dark
matter effects can be modelled by off-diagonal interactions and deformations
of a nontrivial QC like gravitational vacuum structure and analogous scalar
matter fields.

\vskip5pt

\textsf{Keywords:}\ Off--diagonal cosmological metrics; effective
gravitational and scalar field aperiodic structures; Starobinsky-Like
inflation; dark energy and dark matter as quasicrystal structures.
\end{abstract}

\tableofcontents




\section{Introduction}

The Planck data \cite{plank} for modern cosmology prove a remarkable
consistency of Starobinsky $R^{2}$--inflation theory \cite{starob} and a
series of classical works on inflational cosmology \cite%
{much,guth,linde,albrecht}. For reviews of results and changing of modern
cosmology paradigms with dark energy and dark matter, and on modified
gravity theories, MGTs, we cite \cite%
{acelun,nojod1,capoz,clifton,gorbunov,linde2,bambaodin,sami,mimet2,odints2,guend1,guend2,stavr1,stavr2}%
. Various cosmological scenarios studied in the framework of MGTs involve
certain inhomogeneous and local anisotropic vacuum and non--vacuum
gravitational configurations determined by corresponding type (effective)
inflation potentials. In order to investigate such theoretical models, there
are applied advanced numeric, analytic and geometric techniques which allow
us to find exact, parametric and approximate solutions for various classes
of nonlinear systems of partial differential equations, PDEs, considered in
mathematical gravity and cosmology. The main goal of this paper is to
elaborate on geometric methods in acceleration cosmology physics and certain
models with effective quasicrystal like gravitational and scalar field
structures.

The anholonomic frame deformation method, AFDM, (see review and applications
in four dimensional, 4-d, and extra dimension gravity in \cite%
{gvvafdm,vpopa,vsolit2}) allows to construct in certain general forms
various classes of (non) vacuum generic off--diagonal solutions in Einstein
gravity and MGTs. Such solutions may describe cosmological observation data
and explain and predict various types of gravitational and particle physics
effects \cite{cosmv1,cosmv2,eegrg,cosmv3}. Following this geometric method,
we define such nonholonomic frame transformations and deformations of
connections (all determined by the same metric structure) when the
gravitational and matter field equations of motion decouple in general
forms. As a result, we can also integrate in general forms, certain systems
of gravitational and cosmologically important PDEs when the solutions are
determined by generating and integration functions depending (in principle)
on all spacetime coordinates and various classes of integration parameters.

The AFDM is very different from all other former methods applied for
constructing exact solutions, when certain ansatz with higher symmetries
(spherical, cylindrical, certain Lie algebra ones ...) are used for metrics
which can be diagonalized by frame/coordinate transforms. For such
"simplified" ansatz, the gravitational motion and/or cosmological equations
transform into nonlinear systems of ordinary differential equations, ODEs,
which can be solved exactly for some special cases. For instance, there are
diagonal ansatz for metrics with dependence on one radial (or time like)
coordinate and respective transforms of PDEs to ODEs in order to construct
black hole (or homogeneous cosmological) solutions, when the physical
effects are computed for integration constants determined by respective
physical constants. The priority of the AFDM is that we can use various
classes of generating and integration functions in order to construct exact
and parametric solutions determined by nonholonomic nonlinear constraints
and transforms, with generic off-diagonal metrics, generalized connections
and various effective gravitational and matter field configurations. Having
constructed a class of general solutions, we can always impose additional
constraints at the end (for instance, zero torsion, necessary boundary and
symmetry conditions, or to consider homogeneous and isotropic configurations
with nontrivial topology etc.) and search for possible limits to well known
physically important solutions. Here we note that in attempting to find
exact solutions for nonlinear systems we lose a lot of physically important
classes of solutions if we make certain approximations with "simplified"
ansatz from the very beginning. Applying the AFDM, we can generate certain
very general families of exact/parametric solutions after which there are
possibilities to consider additional nonholonomic (non integrable)
constraints and necessary type symmetry/ topology / boundary / asymptotic
conditions which allow us to explain and predict certain observational
and/or experimental data.

Modern cosmological acceleration data \cite{plank} show the existence of
certain complex network filament and aperiodic type structures (with various
fractal like, diffusion processes, nonlinear wave interactions etc.)
determined by dark energy of the Universe and distribution of dark matter,
with "hidden" frame support for respective meta--galactic and galactic
configurations. Such configurations can be modelled by numeric and/or
analytic geometric methods as deformations of an effective quasicrystal, QC,
structure for the gravitational and fundamental scalar fields. Similar ideas
were proposed many years ago in connection both with inflationary cosmology
and quasicrystal physics \cite{steinh1,steinh2,steinh3,albrecht}. On early
works and modern approaches to QC mathematics and physics, we cite \cite%
{emmr,subr,steinh4,penrose1,penrose2,achim,barkan,klee1,klee2} and
references therein. Here we note that it is not possible to introduce
directly a QC like locally anisotropic and inhomogeneous structure described
by solutions of (modified) Einstein equations if we restrict the geometric
approach only for cosmological models based only on the Friedmann--Lema\^{\i}%
tre-Robertson-Worker, FLRW, metric. The geometric objects and physical
values with homogeneous and isotropic metrics are determined by solutions
with integration constants. This does not allow us to elaborate on
realistic, physically motivated cosmological configurations with QC
nontrivial vacuum and non--vacuum structures. Realistic QC like aperiodic
structures can not be described solely via integration constants or certain
Bianchi / Killing type and Lie group symmetries with structure constants.
Solutions with nontrivial gravitational vacuum structure and respective
cosmological scenarios can be generated by prescribing, via generating
functions and generating sources, possible observational QC configurations
following the AFDM as in \cite{cosmv1,cosmv2,eegrg,cosmv3}. In nonholonomic
variables, we can describe the formation and evolution of QC structures as
generalized geometric flow effects and time quasicrystal structures, see
partner works \cite{partner2,partner3} as recent developments and
applications in physics of R. Hamilton and G. Perelman's theory of Ricci
flows \cite{vanp,vrajssrf}. Here, we emphasize that Lyapunov type
functionals (for free energy) are used both in QC and Ricci flow evolution
theories. For geometric evolution theories, such general entropy type
functionals are known as Perelman's functionals with associated
thermodynamic variables. The purpose of the current planned series of papers
is to study generic off--diagonal cosmological solutions with aperiodic
order for (modified) Ricci solitons. In explicit form, the main goal of this
article is to provide a geometric proof that aperiodic QC structures of
vacuum and nonvacuum solutions of gravitational and scalar matter field
equations MGTs and GR result in cosmological solutions mimicking Starobinsky
like inflations and dark energy and dark matter scenarios which are
compatible with both the accelerating cosmology paradigm and observational
cosmological data. We also follow certain methods for the mathematics of
aperiodic order structures summarized in \cite{aperqc}.

This paper is structured as follows. Section \ref{sec2} is devoted to
geometric preliminaries and the main formulas for generating generic
off--diagonal cosmological solutions in MGTs and GR (for proofs and details,
readers are directed to review papers \cite{gvvafdm,cosmv1,cosmv2,eegrg}).
In section \ref{sec3}, we elaborate on methods of constructing exact
solutions with gravitational and scalar field effective QC structure. We
study how aperiodic structures can be defined by generating functions and
matter field sources adapted to off-diagonal gravitational and matter field
interactions and evolution processes. Then, in section \ref{sec4}, we prove
that Starobinsky like inflation scenarios can be determined by effective QC
gravitational and adapted scalar field configurations. In that section we
also outline a formalism for reconstructing analogous QC cosmological
structures, with dark energy and dark matter effects. In section \ref{sec5},
we study certain MGTs configurations when dark energy and dark matter
physics is modelled by gravitational QCs and effective scalar field
configurations. Finally, conclusions and discussion are provided in section %
\ref{sec6}.

\section{Generating Aperiodic Cosmological Solutions in $R^2$ Gravity}

\label{sec2} Let us outline a geometric approach for constructing exact
solutions, (for the purposes of this paper), with aperiodic continuous
and/or discrete, in general, inhomogeneous and locally anisotropic
cosmological structures in (modified) gravity theories. We consider a
general off--diagonal cosmological metrics $\mathbf{g}$ on a four
dimensional, 4d, pseudo--Riemannian manifold $\mathbf{V}$ which can be
parameterized via certain frame transforms as a distinguished metric,
d--metric, in the form
\begin{eqnarray}
\mathbf{g} &=&g_{\alpha \beta }(u)\mathbf{e}^{\alpha }\otimes \mathbf{e}%
^{\beta }=g_{i}(x^{k})dx^{i}\otimes dx^{i}+g_{a}(x^{k},y^{b})\mathbf{e}%
^{a}\otimes \mathbf{e}^{b}  \label{odans} \\
&=&g_{\alpha \beta }(u)\mathbf{e}^{\alpha }\otimes \mathbf{e}^{\beta
},g_{\alpha ^{\prime }\beta ^{\prime }}(u)=g_{\alpha \beta }\mathbf{e}_{\
\alpha ^{\prime }}^{\alpha }\mathbf{e}_{\ \beta ^{\prime }}^{\beta },  \notag
\\
\mathbf{e}^{a} &=&dy^{a}+N_{i}^{a}(u^{\gamma })dx^{i}\mbox{ and }\mathbf{e}%
^{\alpha }=\mathbf{e}_{\ \alpha ^{\prime }}^{\alpha }(u)du^{\alpha ^{\prime
}}.  \label{nbdual}
\end{eqnarray}%
In these formulas, the local coordinates $u^{\gamma }=(x^{k},y^{c}),$ or $%
u=(x,y),$ where indices run respective values $i,j,k,...=1,2$ and $%
a,b,c,...=3,4$ are for conventional 2+2 splitting on$\mathbf{V}$ of
signature $(+++-),$ when $u^{4}=y^{4}=t$ is a time like coordinate and $u^{%
\grave{\imath}}=(x^{i},y^{3})$ are spacelike coordinates with $\grave{\imath}%
,\grave{j},\grave{k},...=1,2,3.$ The values $\mathbf{N}=\{N_{i}^{a}%
\}=N_{i}^{a}$ define a N--adapted decomposition of the tangent bundle $T%
\mathbf{V}=hT\mathbf{V}\oplus vT\mathbf{V}$ into conventional horizontal, h,
and vertical, v, subspaces\footnote{%
If it will not be a contrary statement for an explicit formula, we shall use
the Einstein rule on summation on "up-low" cross indices. Such a system of
"N--adapted notations" with boldface symbols for a nontrivial nonlinear
connections, N--connection structure determined by $\mathbf{N}$ and related
nonholonomic differential geometry is explained in detail in \cite%
{gvvafdm,cosmv1,cosmv2,eegrg,cosmv3} and references therein. We omit such
considerations in this work.}. Such a geometric splitting is nonholonomic
because the basis $\mathbf{e}^{\alpha }=(x^{i},\mathbf{e}^{a})$ is dual to $%
\mathbf{e}_{\alpha }=(\mathbf{e}_{i},e_{a})$
\begin{equation*}
e_{i}=\partial /\partial x^{i}-N_{i}^{a}(u)\partial /\partial
y^{a},e_{a}=\partial _{a}=\partial /\partial y^{a},
\end{equation*}%
which is nonholonomic (equivalently, non-intergrable, or anholonomic) if the
commutators
\begin{equation*}
\mathbf{e}_{[\alpha }\mathbf{e}_{\beta ]}:=\mathbf{e}_{\alpha }\mathbf{e}%
_{\beta }-\mathbf{e}_{\beta }\mathbf{e}_{\alpha }=C_{\alpha \beta }^{\gamma
}(u)\mathbf{e}_{\gamma }
\end{equation*}%
contain nontrivial anholonomy coefficients $C_{\alpha \beta }^{\gamma
}=\{C_{ia}^{b}=\partial _{a}N_{i}^{b},C_{ji}^{a}=\mathbf{e}_{j}N_{i}^{a}-%
\mathbf{e}_{i}N_{j}^{a}\}.$ If such coefficients are nontrivial, a
N--adapted metric (\ref{odans}) can not be diagonalized in a local finite,
or infinite, spacetime region with respect to coordinate frames. Such
metrics are generally off-diagonal and characterized by six independent
nontrivial coefficients from a set $\mathbf{g}=\{g_{\alpha \beta }(u)\}.$ A
frame is holonomic if all corresponding anholonomy coefficients are zero
(for instance, the coordinate frames).

On $\mathbf{V,}$ we can consider a distinguished connection, d--connection, $%
\mathbf{D,}$ structure as a metric--affine (linear) connection preserving
the $N$--connection splitting under parallel transports, i.e. $\mathbf{D}%
=(hD,vD).$ We denote the torsion of $\mathbf{D}$ as $\mathcal{T}=\{\mathbf{T}%
_{\beta \gamma }^{\alpha }\}$, where the coefficients can be computed in
standard form with respect to any (non) holonomic basis. For instance, the
well known Levi--Civita, LC, connection $\nabla $ is a linear connection but
not a d--connection because it does not preserve, under general
frame/coordinate transforms, a h-v--splitting. Prescribing any d--metric and
N--connection structure, we can work on $\mathbf{V}$ in equivalent form with
two different linear connections:
\begin{equation}
(\mathbf{g,N})\rightarrow \left\{
\begin{array}{cc}
\mathbf{\nabla :} & \mathbf{\nabla g}=0;\ ^{\nabla }\mathcal{T}=0,%
\mbox{\
for  the LC--connection } \\
\widehat{\mathbf{D}}: & \widehat{\mathbf{D}}\mathbf{g}=0;\ h\widehat{%
\mathcal{T}}=0,v\widehat{\mathcal{T}}=0,hv\widehat{\mathcal{T}}\neq 0,%
\mbox{
for the canonical d--connection  }.%
\end{array}%
\right.  \label{twocon}
\end{equation}%
In this formula, the $\widehat{\mathbf{D}}=h\widehat{\mathbf{D}}+v\widehat{%
\mathbf{D}}$ is completely defined by $\mathbf{g}$ for any prescribed
N--connection structure $\mathbf{N}$. There is a canonical distortion
relation
\begin{equation}
\widehat{\mathbf{D}}=\nabla +\widehat{\mathbf{Z}}.  \label{candistr}
\end{equation}%
The distortion distinguished tensor, d-tensor, $\widehat{\mathbf{Z}}=\{%
\widehat{\mathbf{Z}}_{\ \beta \gamma }^{\alpha }[\widehat{\mathbf{T}}_{\
\beta \gamma }^{\alpha }]\},$ is an algebraic combination of the
coefficients of the corresponding torsion d-tensor $\widehat{\mathcal{T}}=\{%
\widehat{\mathbf{T}}_{\ \beta \gamma }^{\alpha }\}$ of $\widehat{\mathbf{D}}%
. $ All such values are completely defined by data $(\mathbf{g,N})$ being
adapted to the N--splitting. It should be noted that $\widehat{\mathcal{T}}$
\ is a nonholonomically induced torsion determined by $(C_{\alpha \beta
}^{\gamma },\partial _{\beta }N_{i}^{a},g_{\alpha \beta }).$ It is different
from that considered, for instance, in the Einstein--Cartan, or string
theory, where they consider additional field equations for torsion fields.
We can redefine all geometric constructions for $\widehat{\mathbf{D}}$ in
holonomic or nonholonomic variables for $\nabla $ when the torsion vanishes
in result of nonholonomic deformations.\footnote{%
Using a standard geometric techniques, the torsions, $\widehat{\mathcal{T}}$
and $\ ^{\nabla }\mathcal{T}=0,$ and curvatures, $\widehat{\mathcal{R}}=\{%
\widehat{\mathbf{R}}_{\ \beta \gamma \delta }^{\alpha }\}$ and $\ ^{\nabla }%
\mathcal{R}=\{R_{\ \beta \gamma \delta }^{\alpha }\}$ (respectively, for $%
\widehat{\mathbf{D}}$ and $\nabla )$ are defined and can be computed in
coordinate free and/or coefficient forms. We can define the corresponding
Ricci tensors, $\ \widehat{\mathcal{R}}ic=\{\widehat{\mathbf{R}}_{\ \beta
\gamma }:=\widehat{\mathbf{R}}_{\ \alpha \beta \gamma }^{\gamma }\}$ and $%
Ric=\{R_{\ \beta \gamma }:=R_{\ \alpha \beta \gamma }^{\gamma }\},$ when the
Ricci d-tensor $\widehat{\mathcal{R}}ic$ is characterized $h$-$v$ N-adapted
coefficients, $\widehat{\mathbf{R}}_{\alpha \beta }=\{\widehat{R}_{ij}:=%
\widehat{R}_{\ ijk}^{k},\ \widehat{R}_{ia}:=-\widehat{R}_{\ ika}^{k},\
\widehat{R}_{ai}:=\widehat{R}_{\ aib}^{b},\ \widehat{R}_{ab}:=\widehat{R}_{\
abc}^{c}\}.$ We can also define two different scalar curvature, $\ R:=%
\mathbf{g}^{\alpha \beta }R_{\alpha \beta }$ and $\ \widehat{\mathbf{R}}:=%
\mathbf{g}^{\alpha \beta }\widehat{\mathbf{R}}_{\alpha \beta }=g^{ij}%
\widehat{R}_{ij}+g^{ab}\widehat{R}_{ab}$. Following the two connection
approach (\ref{twocon}), the (pseudo) Riemannian geometry can be
equivalently described by two different geometric data $\left( \mathbf{%
g,\nabla }\right) $ and $(\mathbf{g,N,}\widehat{\mathbf{D}}).$ Using the
canonical distortion relation (\ref{candistr}), we can compute respective
distortions $\widehat{\mathcal{R}}=\ ^{\nabla }\mathcal{R+}\ ^{\nabla }%
\mathcal{Z}$ and $\widehat{\mathcal{R}}ic=Ric+\widehat{\mathcal{Z}}ic$ and $%
\ ^{\nabla }\mathcal{Z}$ and $\widehat{\mathcal{Z}}ic.$%
\par
Application of such geometric methods in GR and MGTs is motivated by the
fact that various types of gravitational and matter field equations
rewritten in nonholonomic variables $(\mathbf{g,N},\widehat{\mathbf{D}})$
can be decoupled and integrated in certain general forms following the AFDM.
This is not possible if we work from the very beginning with the data $%
\left( \mathbf{g,\nabla }\right) $. Nevertheless, necessary type
LC-configurations can be extracted from certain classes of solutions of
(modified) gravitational field equations if additional conditions resulting
in zero values for the canonical d-torsion, $\widehat{\mathcal{T}}=0,$ are
imposed (considering some limits $\widehat{\mathbf{D}}_{\mid \widehat{%
\mathcal{T}}\rightarrow 0}=\mathbf{\nabla ).}$}

The action, $\mathcal{S}$, for a quadratic gravity model with $\widehat{%
\mathbf{R}}^{2}$ and matter fields with Lagrange density $~^{m}\mathcal{L}(%
\mathbf{g,N,\varphi })$ is postulated in the form
\begin{equation}
\mathcal{S}=M_{P}^{2}\int d^{4}u\sqrt{|\mathbf{g}|}[\widehat{\mathbf{R}}%
^{2}+~^{m}\mathcal{L}],  \label{actqmg}
\end{equation}%
where the Planck mass $M_{P}$ is determined by the gravitational constant.
For simplicity, in this paper we consider actions $~^{m}\mathcal{S}=\int
d^{4}u\sqrt{|\mathbf{g}|}~^{m}\mathcal{L}$ depending only on the
coefficients of a metric field and not on their derivatives. In N--adapted
form, the energy--momentum d--tensor can be computed%
\begin{equation*}
\ ^{m}\mathbf{T}_{\alpha \beta }:=-\frac{2}{\sqrt{|\mathbf{g}_{\mu \nu }|}}%
\frac{\delta (\sqrt{|\mathbf{g}_{\mu \nu }|}\ \ ^{m}\mathcal{L})}{\delta
\mathbf{g}^{\alpha \beta }}=\ ^{m}\mathcal{L}\mathbf{g}^{\alpha \beta }+2%
\frac{\delta (\ ^{m}\mathcal{L})}{\delta \mathbf{g}_{\alpha \beta }}.
\end{equation*}%
In the next sections, we shall choose such dependencies of $~^{m}\mathcal{L}$
on (effective) scalar fields $\mathbf{\varphi }$ which will allow us to
model cosmological scenarios with dark mater and dark energy in MGTs in a
compatible form with nontrivial quasicrystal like gravitational and matter
fields. The action $\mathcal{S}$ (\ref{actqmg}) results in the field
equations
\begin{equation}
\widehat{\mathbf{R}}_{\mu \nu }=\mathbf{\Upsilon }_{\mu \nu },  \label{qgreq}
\end{equation}%
\begin{equation*}
\mbox{ where } \mathbf{\Upsilon }_{\mu \nu }=~^{m}\mathbf{\Upsilon }_{\mu
\nu }+~\widehat{\mathbf{\Upsilon }}_{\mu \nu }, \mbox{ for } \ ^{m}\mathbf{%
\Upsilon }_{\alpha \beta }=\frac{1}{2M_{P}^{2}}\ ^{m}\mathbf{T}_{\alpha
\beta }\mbox{ and }\ \widehat{\mathbf{\Upsilon }}_{\mu \nu }=(\frac{1}{4}%
\widehat{\mathbf{R}}-\frac{\widehat{\square }\ \widehat{\mathbf{R}}}{~%
\widehat{\mathbf{R}}})\mathbf{g}_{\mu \nu }+\frac{\widehat{\mathbf{D}}_{\mu }%
\widehat{\mathbf{D}}_{\nu }\ \widehat{\mathbf{R}}}{\widehat{\mathbf{R}}},
\end{equation*}%
\ and $\widehat{\square }:=\widehat{\mathbf{D}}^{2}=\mathbf{g}^{\mu \nu }%
\widehat{\mathbf{D}}_{\mu }\widehat{\mathbf{D}}_{\nu }.$ For $\widehat{%
\mathbf{D}}_{\mid \widehat{\mathcal{T}}\rightarrow 0}=\mathbf{\nabla ,}$ the
equations can be re-defined via conformal transforms $\ g_{\mu \nu
}\rightarrow \widetilde{g}_{\mu \nu }=g_{\mu \nu }e^{-\ln |1+2\widetilde{%
\mathbf{\varphi }}|},$ for $\sqrt{2/3}\mathbf{\varphi }=\ln |1+2\widetilde{%
\mathbf{\varphi }}|,$ which introduces a specific Lagrange density for
matter into the gravitational equations with effective scalar fields. Such a
construction was used in the Starobinsky modified cosmology model \cite%
{starob}. In N--adapted frames, such a scalar field density can be chosen
\begin{equation}
\ ^{m}\widehat{\mathcal{L}}=-\frac{1}{2}\mathbf{e}_{\mu }\mathbf{\varphi }\
\mathbf{e}^{\mu }\mathbf{\varphi }-\ ^{\mathbf{\varphi }}V(\mathbf{\varphi })
\label{sclagr}
\end{equation}%
resulting in matter field equations%
\begin{equation*}
\widehat{\square }\mathbf{\varphi }+\frac{d\ \ ^{\mathbf{\varphi }}V(\mathbf{%
\varphi })}{d\mathbf{\varphi }}=0.
\end{equation*}%
In the above formula, we consider a nonlinear potential for scalar field $%
\phi $
\begin{eqnarray}
\ ^{\mathbf{\varphi }}V(\mathbf{\varphi }) &=&\varsigma ^{2}(1-e^{-\sqrt{2/3}%
\mathbf{\varphi }})^{2},\varsigma =const,  \label{starobpot} \\
\mbox{ when }\ ^{\mathbf{\varphi }}V(\mathbf{\varphi } &\gg &0)\rightarrow
\varsigma ^{2},\ ^{\mathbf{\varphi }}V(\mathbf{\varphi }=0)=0,\ ^{\mathbf{%
\varphi }}V(\mathbf{\varphi }\ll 0)\sim \varsigma ^{2}e^{-2\sqrt{2/3}\mathbf{%
\varphi }}.  \notag
\end{eqnarray}

In accordance with the aim of this work, we shall study scalar fields
potentials $V(\mathbf{\varphi })$ modified by effective quasicrystal
structures, $\mathbf{\varphi \rightarrow \varphi =\varphi }_{0}\mathbf{+\psi
,}$ where $\mathbf{\psi (}x^{i},y^{3},t)$ with crystal, or QC, like phases
described by periodic or quasi-periodic modulations. Such modifications can
be modelled in dynamical phase field crystal, PFC, like form \cite{cross}.
The corresponding 3-d nonrelativistic dynamics are determined by a Laplace
like operator $\ ^{3}\triangle =(\ ^{3}\nabla )^{2},$ with left label 3. In
N--adapted frames with 3+1 splitting the equations for a local minimum
conserving dynamics,
\begin{equation*}
\partial _{t}\mathbf{\psi =}\ ^{3}\triangle \left[ \frac{\delta F[\mathbf{%
\psi }]}{\delta \mathbf{\psi }}\right] ,
\end{equation*}%
with two length scales $l_{\underline{i}}=2\pi /k_{\underline{i}},$ for $%
\underline{i}=1,2.$ Such a local diffusion process is described by a free
energy functional%
\begin{equation*}
F[\mathbf{\psi }]=\int \sqrt{\mid \ ^{3}g\mid }dx^{1}dx^{2}dy^{3}[\frac{1}{2}%
\mathbf{\psi \{-\epsilon +}\prod\limits_{\underline{i}=1,2}(k_{\underline{i}%
}^{2}+\ ^{3}\triangle )^{2}\mathbf{\}\psi +}\frac{1}{4}\mathbf{\psi }^{4}],
\end{equation*}%
where $\mid \ ^{3}g\mid $ is the determinant of the 3-d space metric and $%
\mathbf{\epsilon }$ is a constant. For simplicity, we restrict our
constructions to only non-relativistic diffusion processes, see Refs. \cite%
{vdif1,vdif2} for relativistic and N--adapted generalizations. $\ $

We are able to explicitly integrate the gravitational field equations (\ref%
{qgreq}) and a d--metric (\ref{actqmg}) for (effective) matter field
configurations parameterized with respect to N--adapted frames in the form
\begin{equation}
\mathbf{\Upsilon }_{\ \nu }^{\mu }=\mathbf{e}_{\ \mu ^{\prime }}^{\mu }%
\mathbf{e}_{\nu }^{\ \nu ^{\prime }}\mathbf{\Upsilon }_{\ \nu ^{\prime
}}^{\mu ^{\prime }}[\ \ ^{m}\mathcal{L(}\mathbf{\varphi }+\mathbf{\psi ),}\
\widehat{\mathbf{\Upsilon }}_{\mu \nu }]=diag[~\ _{h}\Upsilon (x^{i})\delta
_{j}^{i},\Upsilon (x^{i},t)\delta _{b}^{a}],  \label{paramsourc}
\end{equation}%
for certain vielbein transforms $\mathbf{e}_{\ \mu ^{\prime }}^{\mu
}(u^{\gamma })$ and their duals $\mathbf{e}_{\nu }^{\ \nu ^{\prime
}}(u^{\gamma }),$ when $\mathbf{e}_{\ }^{\mu }=\mathbf{e}_{\ \mu ^{\prime
}}^{\mu }du^{\mu ^{\prime }},$ and $\mathbf{\Upsilon }_{\ \nu ^{\prime
}}^{\mu ^{\prime }}=~^{m}\mathbf{\Upsilon }_{\ \nu ^{\prime }}^{\mu ^{\prime
}}+~\widehat{\mathbf{\Upsilon }}_{\ \nu ^{\prime }}^{\mu ^{\prime }}.$ The
values $~\ _{h}\Upsilon (x^{i})$ and $\Upsilon (x^{i},t)$ will be considered
as generating functions for (effective) matter sources imposing certain
nonholonomic frame constraints on (effective) dynamics of matter fields.

The system of modified Einstein equations (\ref{qgreq}) with sources (\ref%
{paramsourc}) can be integrated in general form by such an off--diagonal
ansatz (see details in Refs. \cite{gvvafdm,cosmv1,cosmv2,eegrg,cosmv3}):
\begin{eqnarray}
\ g_{i} &=&e^{\ \psi (x^{k})}\mbox{ as a solution of }\psi ^{\bullet \bullet
}+\psi ^{\prime \prime }=2~\ _{h}\Upsilon ;  \label{offdparam} \\
g_{3} &=&\omega ^{2}(x^{i},y^{3},t)h_{3}(x^{i},t)=h_{3}^{[0]}(x^{k})-\frac{1%
}{4}\int dt\frac{\partial _{t}(\Psi ^{2})}{\ \Upsilon };  \notag \\
g_{4} &=&\omega ^{2}(x^{i},y^{3},t)h_{4}(x^{i},t)=-\frac{1}{4}\frac{\partial
_{t}(\Psi ^{2})}{\Upsilon ^{2}}\left( h_{3}^{[0]}(x^{k})-\frac{1}{4}\int dt%
\frac{\partial _{t}(\Psi ^{2})}{4\Upsilon }\right) ^{-1};  \notag \\
N_{k}^{3} &=&n_{k}(x^{i},t)=\ _{1}n_{k}(x^{i})+\ _{2}n_{k}(x^{i})\int dt%
\frac{(\partial _{t}\Psi )^{2}}{\Upsilon ^{2}\left\vert
h_{3}^{[0]}(x^{i})-\int dt\ \partial _{t}(\Psi ^{2})/4\Upsilon \right\vert
^{5/2}};  \notag \\
\ \ N_{i}^{4} &=&w_{i}(x^{k},t)=\partial _{i}\ \Psi /\ \partial _{t}\Psi ;
\notag \\
\ \ \omega  &=&\omega \lbrack \Psi ,\Upsilon ]%
\mbox{ is any solution of the
1st order system }\mathbf{e}_{k}\omega =\partial _{k}\omega +n_{k}\partial
_{3}\omega +w_{k}\partial _{t}\omega =0.  \notag
\end{eqnarray}%
In these formulas, $\Psi (x^{i},t)$ and $\omega (x^{i},y^{3},t)$ are
generating functions; $\ _{h}\Upsilon (x^{i})$ and $\Upsilon (x^{i},t)$ are
respective generating $h$- and $v$--sources; $\ _{1}n_{k}(x^{i}),\
_{2}n_{k}(x^{i})$ and $h_{a}^{[0]}(x^{k})$ are integration functions. Such
values can be defined in explicit form for certain symmetry / boundary /
asymptotic conditions which have to be considered in order to describe
certain observational cosmological data (see next sections). The
coefficients (\ref{offdparam}) generate exact and/or parametric solutions
for any nontrivial $\omega ^{2}=|h_{3}|^{-1}.$ As a particular case, we can
chose $\omega ^{2}=1$ which allows us to construct generic off--diagonal
solutions with Killing symmetry on $\partial _{3}.$

The quadratic elements for such general locally anisotropic and
inhomogeneous cosmological solutions with nonholonomically induced torsion
are parameterized in this form:
\begin{eqnarray}
ds^{2} &=&\ g_{\alpha \beta }(x^{k},y^{3},t)du^{\alpha }du^{\beta }=e^{\
\psi }[(dx^{1})^{2}+(dx^{2})^{2}]+  \label{dmetrgsol} \\
&&\ \ \omega ^{2}\ \{h_{3}[dy^{3}+(\ _{1}n_{k}+\ _{2}n_{k}\int dt\frac{%
(\partial _{t}\Psi )^{2}}{\ \Upsilon ^{2}|h_{3}|^{5/2}})dx^{k}]^{2}-\frac{1}{%
4h_{3}}\left[ \frac{\partial _{t}\Psi }{\Upsilon }\right] ^{2}\ [dt+\frac{%
\partial _{i}\Psi }{\ \partial _{t}\Psi }dx^{i}]^{2}\}.  \notag
\end{eqnarray}%
In principle, we can consider that $h_{3}$ and $\Upsilon $ are certain
generating functions when $\Psi \lbrack h_{3},B,\Upsilon ]$ is computed for $%
\omega ^{2}=1$ from $\partial _{t}(\Psi ^{2})=B(x^{i},t)/\Upsilon $ as a
solution of
\begin{equation*}
\Upsilon \left( h_{3}^{[0]}(x^{k})-\int dtB\right) h_{3}(x^{i},t)=-B.
\end{equation*}%
This equation is equivalent to the second equation (\ref{offdparam}) up to
re-definition of the integration function $h_{3}^{[0]}(x^{k}).$ Various
classes of exact solutions with nontrivial nonholonomically induced torsion
can be constructed. For instance, choosing data $(\Psi ,\Upsilon )$ for
solitonic like functions and/or for various singular, or discrete like
structures. Such generic off--diagonal metrics can encode nontrivial vacuum
and non-vacuum configurations, fractional and diffusion processes, and
describe structure formation for evolving universes, effects with
polarization of gravitational and matter field interaction constants,
modified gravity scenarios etc., see examples in Refs. \cite%
{gvvafdm,cosmv1,cosmv2,eegrg,vdif1,vdif2}.

The class of metrics (\ref{dmetrgsol}) defines exact solutions for the
canonical d--connection $\widehat{\mathbf{D}}$ in $\ \widehat{\mathbf{R}}%
^{2} $ gravity with nonholonomically induced torsion and effective scalar
field encoded into a gravitationally polarized vacuum. We can impose
additional constraints on generating functions and sources in order to
extract Levi--Civita configurations. This is possible for a special class of
generating functions and sources when for $\Psi =\check{\Psi}%
(x^{i},t),\partial _{t}(\partial _{i}\check{\Psi})=\partial _{i}(\partial
_{t}\check{\Psi})$ and $\Upsilon (x^{i},t)=\Upsilon \lbrack \check{\Psi}],$
or $\Upsilon =const.$ For such LC--solutions, we find some functions $\check{%
A}(x^{i},t)$ and $n(x^{k})$ when
\begin{equation*}
w_{i}=\check{w}_{i}=\partial _{i}\check{\Psi}/\partial _{t}\check{\Psi}%
=\partial _{i}\check{A}\mbox{ and }n_{k}=\check{n}_{k}=\partial _{k}n(x^{i}).
\end{equation*}%
The corresponding quadratic line element can be written
\begin{equation}
ds^{2}=\ e^{\ \psi }[(dx^{1})^{2}+(dx^{2})^{2}]+\omega
^{2}\{h_{3}[dy^{3}+(\partial _{k}n)dx^{k}]^{2}-\frac{1}{4h_{3}}\left[ \frac{%
\partial _{t}\check{\Psi}}{\Upsilon }\right] ^{2}\ [dt+(\partial _{i}\check{A%
})dx^{i}]^{2}\}.  \label{lcqel}
\end{equation}%
Both classes of solutions (\ref{dmetrgsol}) and/or (\ref{lcqel}) posses
additional nonlinear symmetries which allow us to redefine the generation
function and generating source in a form determined by an effective (in the $%
v$-subspace) gravitational constant. For certain special parameterizations
of $(\check{\Psi},\Upsilon )$ and other coefficients, we can reproduce
Bianchi like universes, extract FLRW like metrics, or various inhomogeneous
and locally anisotropic configurations in GR. Using generic off-diagonal
gravitational and matter field interactions, we can mimic MGTs effects, or
model fractional/ diffusion / crystal like structure formation. Finally, we
note that, such metrics (\ref{lcqel}) can not be localized in finite or
infinite spacetime regions if there are nontrivial anholonomy coefficients $%
C_{\alpha \beta }^{\gamma }.$

\section{Modified Gravity with Quasicrystal Like Structures}

\label{sec3}To introduce thermodynamical like characteristics for
gravitational and scalar fields we consider an additional 3+1 splitting when
the off--diagonal metric ansatz of type (\ref{odans}), (\ref{dmetrgsol}) (%
\ref{lcqel}) can be re-written in the form
\begin{eqnarray}
\mathbf{g} &=&b_{i}(x^{k})dx^{i}\otimes dx^{i}+b_{3}(x^{k},y^{3},t)\mathbf{e}%
^{3}\otimes \mathbf{e}^{3}-\breve{N}^{2}(x^{k},y^{3},t)\mathbf{e}^{4}\otimes
\mathbf{e}^{4},  \label{lapsnonh} \\
\mathbf{e}^{3} &=&dy^{3}+n_{i}(x^{k},t)dx^{i},\,\,\,\,\mathbf{e}%
^{4}=dt+w_{i}(x^{k},t)dx^{i}.  \notag
\end{eqnarray}%
In such a case, the 4--d metric $\mathbf{g}$ is considered as an extension
of a 3--d metric $b_{\grave{\imath}\grave{j}}=diag(b_{\grave{\imath}%
})=(b_{i},b_{3})$ on a family of 3-d hypersurfaces $\widehat{\Xi }_{t}$
parameterized by $t.$ \ We have
\begin{equation}
b_{3}=g_{3}=\omega ^{2}h_{3}\mbox{ and }\breve{N}^{2}(u)=-\omega
^{2}h_{4}=-g_{4},  \label{shift1}
\end{equation}%
defining a lapse function $\breve{N}(u).$ For such a double 2+2 and 3+1
fibration, $\widehat{\mathbf{D}}=(\widehat{D}_{i},\widehat{D}_{a})=(\widehat{%
D}_{\grave{\imath}},\widehat{D}_{t})$ (in coordinate free form, we write $(\
^{q}\widehat{D},\ ^{t}\widehat{D})$). Similar splitting can be performed for
the LC-operator, $\nabla =(\nabla _{i},\nabla _{a})=(\nabla _{\grave{\imath}%
},\nabla _{t})=(\ ^{q}\nabla ,\ ^{t}\nabla ).$ For simplicity, we elaborate
the constructions for solutions with Killing symmetry on $\partial _{3}.$

\subsection{Generating functions with 3d quasicristal like structure}

Gravitational QC like structures can be defined by generic off--diagonal
exact solutions if we choose a generating function $\Psi =\Phi $ as a
solution of an evolution equation with conserved dynamics of type
\begin{equation}
\frac{\partial \Phi }{\partial t}=\ ^{b}\widehat{\Delta }\left[ \frac{\delta
F}{\delta \Phi }\right] =-\ ^{b}\widehat{\Delta }(\Theta \Phi +Q\Phi
^{2}-\Phi ^{3}),  \label{evoleq}
\end{equation}%
where the canonically nonholonomically deformed Laplace operator$\ ^{b}%
\widehat{\Delta }:=(\ ^{b}\widehat{D})^{2}=q^{\grave{\imath}\grave{j}}%
\widehat{D}_{\grave{\imath}}\widehat{D}_{\grave{j}}$ as a distortion of $\
^{b}\Delta :=(\ ^{b}\nabla )^{2}$ can be defined on any $\widehat{\Xi }_{t}.$
Such distortions of differential operators can be always computed using
formulas (\ref{candistr}). The functional $F$ in the evolution equation (\ref%
{evoleq}) defines an effective free energy (it can be associate to a model
of dark energy, DE)
\begin{equation}
F[\Phi ]=\int \left[ -\frac{1}{2}\Phi \Theta \Phi -\frac{Q}{3}\Phi ^{3}+%
\frac{1}{4}\Phi ^{4}\right] \sqrt{b}dx^{1}dx^{2}\delta y^{3},  \label{dener}
\end{equation}%
where $b=\det |b_{\grave{\imath}\grave{j}}|,\delta y^{3}=\mathbf{e}^{3}$ and
the operator $\Theta $ and parameter $Q$ will be defined below. Such a
configuration is stabilized nonlinearly by the cubic term when the second
order resonant interactions are varied by setting the value of $Q.$ The
average value $\overline{\Phi }$ of the generating function $\Phi $ is
conserved for any fixed $t.$ This means that $\overline{\Phi }$ can be
thought of as an effective parameter of the system and that we can choose $%
\overline{\Phi }_{|t=t_{0}}=0$ since other values can be redefined and
accommodated by altering $\Theta $ and $Q.$ Further evolution can be
computed for any solution of type (\ref{dmetrgsol}) and/or (\ref{lcqel}).

The effective free energy $F[\Phi ]$ defines an analogous 3-d phase
gravitational field crystal (APGFC) model that generates modulations with
two length scales for off--diagonal cosmological configurations. This model
consists of a nonlinear PDE with conserved dynamics. It describes (in
general, relativistic) time evolution of $\Phi $ over diffusive time scales.
For instance, we can elaborate such a APGFC model in a form including
resonant interactions that may \ occur in the case of icosahedral symmetry
considered for standard quasicrystals in \cite{emmr,subr}. In this work,
such gravitational structures will be defined by redefining $\Phi $ into
respective generating functions $\Psi $ or $\check{\Psi}.$ Let us explain
the respective geometric constructions with changing the generating data $%
(\Psi ,\ \Upsilon )\leftrightarrow (\Phi ,\widetilde{\Lambda }=const)$
following the conditions%
\begin{eqnarray}
\frac{\partial _{t}(\Psi ^{2})}{\Upsilon } &=&\frac{\partial _{t}(\Phi ^{2})%
}{\widetilde{\Lambda }},\mbox{ which is equivalent to }  \label{nonltr} \\
\Phi ^{2} &=&\widetilde{\Lambda }\int dt\Upsilon ^{-1}\partial _{t}(\Psi
^{2})\mbox{ and/or }\Psi ^{2}=\widetilde{\Lambda }^{-1}\int dt\Upsilon
\partial _{t}(\Phi ^{2}).  \notag
\end{eqnarray}%
As a result, we can write respective $v$- and $hv$--coefficients in (\ref%
{offdparam}) in terms of $\Phi $ (redefining the integration functions),
\begin{eqnarray}
h_{3}(x^{i},t) &=&h_{3}[\Phi ]=h_{3}^{[0]}(x^{k})-\frac{\Phi ^{2}}{4%
\widetilde{\Lambda }}; \label{dmcoef1}  \\
h_{4}(x^{i},t)&=& -\frac{1}{4}\frac{\partial _{t}(\Phi
^{2})}{\Upsilon \widetilde{\Lambda }}\left( h_{3}^{[0]}(x^{k})-\frac{\Phi
^{2}}{4\widetilde{\Lambda }}\right) ^{-1}=\frac{1}{\Upsilon }\frac{\partial
_{t}(\Phi ^{2})}{\Phi ^{2}-h_{3}^{[0]}(x^{k})};  \notag \\
n_{k} &=&\ _{1}n_{k}+\ _{2}n_{k}\int dt\ \frac{h_{4}[\Phi ]}{|\ h_{3}[\Phi
]|^{3/2}}=\ _{1}n_{k}+\ _{2}n_{k}\int dt\frac{(\partial _{t}\Phi )^{2}}{4|%
\widetilde{\Lambda }\int dt\Upsilon \partial _{t}(\Phi ^{2})|\ |h_{3}[\Phi
]|^{5/2}}\ \mbox{ and } \notag  \\
w_{i} &=&\frac{\partial _{i}\ \Psi }{\partial _{t}\Psi }=\frac{\partial
_{i}\ \Psi ^{2}}{\partial _{t}(\Psi ^{2})}=\frac{\partial _{i}\ \int
dt\Upsilon \ \partial _{t}(\Phi ^{2})}{\Upsilon \partial _{t}(\Phi ^{2})}.
 \notag
\end{eqnarray}

The nonlinear symmetry (\ref{nonltr}) allows us to change generate such
effective sources (\ref{paramsourc}) which allow to generate QC structures
in self-consistent form when
\begin{equation}
\Upsilon (x^{k},t)\rightarrow \ \Lambda =\ \ ^{f}\Lambda +\ ^{\varphi
}\Lambda ,  \label{sourc4d}
\end{equation}%
with associated effective cosmological constants in MGT, $\ \ ^{f}\Lambda ,$
and for the effective QC structure, $\ ^{\varphi }\Lambda .$ We can identify
$\widetilde{\Lambda }$ with $\Lambda ,$ or any other value $\ \ ^{f}\Lambda
, $ or $\ ^{\varphi }\Lambda $ depending on the class of models with
effective gauge interactions we consider in our work.

Let us explain how the formation and stability of gravitational
configurations with icosahedral quasicrystalline structures can be studied
using a dynamical phase field crystal model with evolution equations (\ref%
{evoleq}). Such a 3-d QC structure is stabilized by nonlinear interactions
between density waves at two length scales \cite{subr}. Using a generating
function $\Phi ,$ we elaborate a 3-d effective phase field crystal model
with two length scales as in the so-called Lifshitz--Petrich model \cite%
{lifsh}. The density distribution of matter mimics a "solid" or a "liquid "
on the microscopic length. The role of operator $\Theta $ to allow two wave
marginal numbers and to introduce possible spatio-temporal chaos is
discussed in \cite{ruck,subr}. The effect is similar at metagalactic scales
when $\Phi $ has a two parametric dependence with $k=1$ (the system is
weakly stable) and $k=1/\tau $ (where, for instance, for $\tau =2\cos \frac{%
\pi }{5}=1.6180$ we obtain the golden ratio, when the system is weakly
unstable).

Choosing a QC-type form for $\Phi $ and determining the coefficients of the
d--metric in the form (\ref{dmcoef1}), we generate a QC like structure for
generic off--diagonal gravitational field interactions. Such a structure is
formed by some type ordered arrangements of galaxies (as \ "atoms") with
very rough rotation and translation symmetries. A more realistic picture of
the observational data for the Universe is for a non crystal structure with
lack of the translational symmetry but yet with certain discrete
observations. There is a certain analogy of such configurations for
quasiperiodic two and three dimensional space like configurations, for
instance, in metallic alloys, or nanoparticles, [as a review, see \cite%
{ruck,subr,lifsh} and references therein] and at meta-galactic scales when
the nontrivial vacuum gravitational cosmological structure is generated as
we consider in this section.

\subsection{Effective scalar fields with quasicrystal like structure}

Following our system of notations, we shall put a left label "q" to the
symbols for geometric/ physical objects in order to emphasize that they
encode an aperiodic QC geometric structure and write, for instance, $\left(
\ \ ^{q}\mathbf{g},\ \ ^{q}\widehat{\mathbf{D}},\ \ ^{q}\varphi \right) .$
We shall omit left labels for continuous configurations and/or if that will
simplify notations, while simultaneously not resulting in ambiguities.

The quadratic gravity theory with action (\ref{actqmg}) is invariant (both
for $\nabla $ and $\widehat{\mathbf{D}}$) under global dilatation symmetry
with a constant $\sigma ,$
\begin{equation}
g_{\mu \nu }\rightarrow e^{-2\sigma }g_{\mu \nu },\varphi \rightarrow
e^{2\sigma }\widetilde{\varphi }.  \label{dilatonsym}
\end{equation}%
We can pass from the Jordan to the Einstein frame with a redefinition $%
\varphi =\sqrt{3/2}\ln |2\widetilde{\varphi }|$ and obtain
\begin{equation}
\ ^{\Phi }\mathcal{S}=\int d^{4}u\sqrt{|g|}\left( \frac{1}{2}\widehat{%
\mathbf{R}}-\frac{1}{2}\mathbf{e}_{\mu }\varphi \ \mathbf{e}^{\mu }\varphi
-2\Lambda \right) ,  \label{linearact}
\end{equation}%
where the scalar potential $\ ^{\varphi }V(\varphi )$ in (\ref{starobpot})
is transformed into an effective cosmological constant term $\Lambda $ using
$(\Psi ,\ \Upsilon )\leftrightarrow (\Phi ,\widetilde{\Lambda })$ (\ref%
{nonltr}). Such an integration constant can be positive / negative / zero,
respectively for de Sitter / anti de Sitter / flat space.

The corresponding field equations derived from $\ ^{E}\mathcal{S}$ are
\begin{eqnarray}
\widehat{\mathbf{R}}_{\mu \nu }-\mathbf{e}_{\mu }\varphi \ \mathbf{e}_{\nu
}\varphi -2\Lambda \mathbf{g}_{\mu \nu } &=&0,  \label{starobh} \\
\widehat{\mathbf{D}}^{2}\varphi &=&0.  \label{scfeq}
\end{eqnarray}%
We obtain a theory with an effective scalar field adapted to a nontrivial
vacuum QC structure encoded into $\mathbf{g}_{\mu \nu },\mathbf{e}_{\mu }$
and $\widehat{\mathbf{D}}$ as generic off--diagonal cosmological solutions.
At the end of this section, we consider three examples of such QC
gravitational-scalar field configurations as aperiodic and mixed continuous
and discrete solutions of the gravitational and matter field equations (\ref%
{starobh}) and (\ref{scfeq}).

\subsubsection{Scalar field N--adapted to gravitational quasicrystals}

In order to generate integrable off--diagonal solutions, we consider certain
special conditions for the effective scalar field $\varphi $ when $\mathbf{e}%
_{\alpha }\varphi =\ ^{0}\varphi _{\alpha }=const$ in N-adapted frames. For
such configurations, $\widehat{\mathbf{D}}^{2}\varphi =0.$ We restrict our
models to configurations of $\phi ,$ which can be encoded into N--connection
coefficients
\begin{equation}
\mathbf{e}_{i}\varphi =\partial _{i}\varphi -n_{i}\partial _{3}\varphi
-w_{i}\partial _{t}\varphi =\ ^{0}\varphi _{i};\,\,\,\partial _{3}\varphi =\
^{0}\varphi _{3};\,\,\,\,\partial _{t}\varphi =\ ^{0}\varphi _{4};\,\,\,\,%
\mbox{ for }\ ^{0}\varphi _{1}=\ ^{0}\varphi _{2}\mbox{ and }\ ^{0}\varphi
_{3}=\ ^{0}\varphi _{4}.  \label{scf}
\end{equation}%
This way we encode the contribution of scalar field configurations into
additional source
\begin{equation*}
\ ^{\varphi }\widetilde{\Upsilon }=~\ ^{\varphi }\widetilde{\Lambda }%
_{0}=const\ \mbox{ and }\ ^{\varphi }\Upsilon =\ ^{\varphi }\Lambda
_{0}=const
\end{equation*}%
even the gravitational vacuum structure is a QC modeled by $\Phi $ as a
solution of (\ref{evoleq}).

\subsubsection{Scalar and rescaled QC generating functions}

The scalar field equations (\ref{evoleq}) can be solved if $\varphi =Z\Phi ,$
for $Z=const\neq 0.$ The conditions (\ref{scf}) with $\ ^{0}\varphi _{1}=\
^{0}\varphi _{2}=\ ^{0}\varphi _{3}=0$ and nontrivial $\widehat{\Gamma }%
_{44}^{4}=-\partial _{t}h_{4}/h_{4}$ transform into
\begin{eqnarray}
\,\partial _{t}\varphi &=&-\ ^{b}\widehat{\Delta }(\Theta \varphi +Q\varphi
^{2}-\varphi ^{3}),  \label{shoh} \\
\partial _{i}\varphi -w_{i}\partial _{t}\varphi &=&\partial _{i}\varphi -%
\frac{\partial _{i}\Phi }{\partial _{t}\Phi }\partial _{t}\varphi \equiv 0,
\notag \\
\widehat{\mathbf{D}}^{2}\varphi &=&h_{4}^{-1}(1+\widehat{\Gamma }%
_{44}^{4})\partial _{t}\varphi =0.  \label{shoh3}
\end{eqnarray}%
For $h_{4}(x^{k},t)$ given by (\ref{dmcoef1}), we obtain nontrivial
solutions of (\ref{shoh3}) if $1+\widehat{\Gamma }_{44}^{4}=0.$ This
additionally constrains $\Phi ,$ i.e. $\varphi =Z\Phi ,$ to the condition $%
2\partial _{t}\varphi =\frac{4h_{4}^{[0]}(x^{k})}{\widetilde{\Lambda }%
Z^{2}\varphi }-\varphi $. Together with (\ref{shoh}) we obtain that
N--adapted scalar fields mimic a QC structure if
\begin{equation*}
\widetilde{\Lambda }Z^{2}\varphi \left[ \varphi -\ ^{b}\widehat{\Delta }%
(\Theta \varphi +Q\varphi ^{2}-\varphi ^{3})\right] =2h_{4}^{[0]}(x^{k}).
\end{equation*}%
Using different scales, we can consider the energy of such QC scalar
structures as hidden energies for dark matter, DM, modeled by $\varphi ,$
determined by an effective functional
\begin{equation}
\ ^{DM}F[\varphi ]=\int \left[ -\frac{1}{2}\varphi \widehat{\Theta }\varphi -%
\frac{\widehat{Q}}{3}\varphi ^{3}+\frac{1}{4}\varphi ^{4}\right] \sqrt{b}%
dx^{1}dx^{2}\delta y^{3},  \label{dmat}
\end{equation}%
where operators $\widehat{\Theta }$ and $\widehat{Q}$ have to be chosen in
some forms compatible to observational data for the standard matter
interacting with the DM. Even the QC structures for the gravitational fields
(with QC configurations for the dark energy, DE) and for the DM can be
different, we parameterize $\ F$ and $\ ^{DM}F$ in similar forms because
such values are described effectively as exact solutions of Starobinsky like
model with quadratic Ricci scalar term. Here we note that such a similar $%
\varphi $--model was studied with a similar\ Lyapunov functional (effective
free energy)$\ ^{DM}F[\varphi ]$ resulting in the Swift--Hohenberg equation (%
\ref{shoh}), see details in Refs. \cite{shoh,lifsh}.

\section{Aperiodic QC Starobinsky Like Inflation}

\label{sec4}The Starobinsky model described an inflationary de Sitter
cosmological solution by postulating a quadratic on Ricci scalar action \cite%
{starob}. In nonholonomic variables, such MGTs were developed in \cite%
{cosmv1,cosmv2,eegrg,cosmv3}.

\subsection{Inflation parameters determined by QC like structures}

Although the Starobinsky cosmological model might appear not to involve any
quasicrystal structure as we described in previous section, it is in fact
conformally equivalent to a nonholonomic deformation of the Einstein gravity
coupled to an effective QC structure that may drive inflation and
acceleration scenarios. This follows from the fact that we can linearize the
$\widehat{\mathbf{R}}^{2}$--term in (\ref{actqmg}) as we considered for the
action (\ref{linearact}). Let us introduce an auxiliary Lagrange field $%
\lambda (u)$ for a constant $\varsigma =8\pi /3\mathcal{M}^{2}$ for a
constant $\mathcal{M}$ of mass dimension one, with $\kappa ^{2}=8\pi G$ for
the Newton's gravitational constant $G=1/M_{P}^{2}$ and Planck's mass, and
perform respective conformal transforms with dilaton symmetry (\ref%
{dilatonsym}). We obtain that the action for our MGT can be written in three
equivalent forms,%
\begin{eqnarray}
\mathcal{S} &=&\frac{1}{2\kappa ^{2}}\int d^{4}u\sqrt{|\mathbf{g}|}\left\{
\widehat{\mathbf{R}}[\mathbf{g}]+\varsigma \widehat{\mathbf{R}}^{2}[\mathbf{g%
}]\right\} ,\mbox{ with }\left\{
\begin{array}{c}
\mathbf{g}_{\mu \nu }\rightarrow \widetilde{\mathbf{g}}_{\mu \nu
}=[1+2\varsigma \lambda (u)]\mathbf{g}_{\mu \nu } \\
\lambda (u)\rightarrow \varphi (u):=\sqrt{3/2}\ln [1+2\varsigma \lambda (u)]%
\end{array}%
\right. ,  \notag \\
&\rightarrowtail &\frac{1}{2\kappa ^{2}}\int d^{4}u\sqrt{|\mathbf{g}|}%
\left\{ [1+2\varsigma \lambda (u)]\widehat{\mathbf{R}}[\mathbf{g}]-\varsigma
\lambda ^{2}(u)\right\}  \notag \\
&\rightarrowtail &\frac{1}{2\kappa ^{2}}\int d^{4}u\sqrt{|\widetilde{\mathbf{%
g}}|}\left\{ \widehat{\mathbf{R}}[\widetilde{\mathbf{g}}]-\frac{1}{2}%
\widetilde{\mathbf{g}}_{\mu \nu }\mathbf{e}^{\nu }\varphi \ \mathbf{e}^{\mu
}\varphi -\ ^{\mathbf{\varphi }}V(\mathbf{\varphi })\right\} ,
\label{mgract}
\end{eqnarray}%
with effective potential $\ ^{\mathbf{\varphi }}V(\mathbf{\varphi })$ (\ref%
{starobpot}) with for a gravitationally modified QC structure $\mathbf{%
\varphi }$--field which for $(\Psi ,\ \Upsilon )\leftrightarrow (\Phi ,%
\widetilde{\Lambda })$ (\ref{nonltr}) defines N--adapted configurations of
type (\ref{scf}) or (\ref{shoh}). Such nonlinear transforms are possible
only for generic off-diagonal cosmological solutions constructed using the
AFDM. We shall write $\ ^{\mathbf{q}}V(\ ^{\mathbf{q}}\mathbf{\varphi })$
for certain effective scalar like structures determined by a nontrivial QC
configuration with $\mathbf{g}_{\mu \nu }\rightarrow \widetilde{\mathbf{g}}%
_{\mu \nu }$ and $\mathbf{\varphi =\ ^{\mathbf{q}}\varphi }$ described above.

In order to understand how actions of type (\ref{mgract}) with effective
free energy $F$ (\ref{dener}), for DE, and $^{DM}F$ (\ref{dmat}), for DM,
encode conditions for inflation like in the Starobinsky quadratic gravity,
let us consider small off--diagonal deformations of FLRW metrics to
solutions of type\textit{\ }(\ref{dmetrgsol}) and (\ref{lcqel}). We
introduce a new time like coordinate $\widehat{t},$ when $t=t(x^{i},\widehat{%
t})$ and $\sqrt{|h_{4}|}\partial t/\partial \widehat{t}$, and a scale factor
$\widehat{a}(x^{i},\widehat{t})$ when the d--metric (\ref{odans}) can be
represented in the form%
\begin{eqnarray}
ds^{2} &=&\widehat{a}^{2}(x^{i},\widehat{t})[\eta _{i}(x^{k},\widehat{t}%
)(dx^{i})^{2}+\widehat{h}_{3}(x^{k},\widehat{t})(\mathbf{e}^{3})^{2}-(%
\widehat{\mathbf{e}}^{4})^{2}],  \label{defdm} \\
\mbox{ where }\eta _{i} &=&\widehat{a}^{-2}e^{\psi },\widehat{a}^{2}\widehat{%
h}_{3}=h_{3},\mathbf{e}^{3}=dy^{3}+\partial _{k}n~dx^{k},\widehat{\mathbf{e}}%
^{4}=d\widehat{t}+\sqrt{|h_{4}|}(\partial _{i}t+w_{i}).  \notag
\end{eqnarray}%
Using a small parameter $\varepsilon ,$ with $0\leq \varepsilon <1,$ we
model off--diagonal deformations if
\begin{equation}
\eta _{i}\simeq 1+\varepsilon \chi _{i}(x^{k},\widehat{t}),\partial
_{k}n\simeq \varepsilon \widehat{n}_{i}(x^{k}),\sqrt{|h_{4}|}(w_{i})\simeq
\varepsilon \widehat{w}_{i}(x^{k},\widehat{t}).  \label{smalld}
\end{equation}%
This corresponds to a subclass of generating functions, which for $%
\varepsilon \rightarrow 0$, result in $\Psi (t),$ or $\check{\Psi}(t),$ and,
correspondingly $\Phi (t),$ and generating source $\Upsilon (t)$ in a form
compatible to $\widehat{a}(x^{i},\widehat{t})\rightarrow $ $\widehat{a}(t),%
\widehat{h}_{3}(x^{i},\widehat{t})\rightarrow \widehat{h}_{3}(\widehat{t})$
etc. Conditions of type (\ref{smalld}) and homogeneous limits for generating
functions and sources have to be imposed after a locally anisotropic
solution (for instance, of type (\ref{lcqel})), was constructed in explicit
form. If we impose homogeneous conditions from the very beginning, we
transform the (modified) Einstein equations with scalar filed in a nonlinear
system of ODEs which do not describe gravitational and scalar field
analogous quasicrystal structures. Applying the AFDM with generating and
integration functions we directly solve nonlinear systems of PDEs and new
classes of cosmological solutions are generated even in diagonal limits
because of generic nonlinear and nonholonomic character of off--diagonal
systems in MGFT. For $\varepsilon \rightarrow 0$ and $\widehat{a}(x^{i},%
\widehat{t})\rightarrow $ $\widehat{a}(t),$ we obtain scaling factors which
are very different from those in the FLRW cosmology with GR solutions.
Nevertheless, we can mimic such cosmological models using redefined
parameters and possible small off--diagonal deformations of cosmological
evolution for MGTs as we explain in details in \cite%
{cosmv1,cosmv2,eegrg,cosmv3}. In this work, we consider effective sources
encoding contributions from the QC gravitational and scalar field
structures, with
\begin{equation*}
\widehat{a}^{2}\widehat{h}_{3}=h_{3}^{[0]}(x^{k})-\Phi ^{2}/4\widetilde{%
\Lambda },
\end{equation*}%
where $\partial _{t}(\Phi ^{2})=\widetilde{\Lambda }$ $\partial _{t}(\Psi
^{2})/\Upsilon ,$ as follows respectively from formulas (\ref{dmcoef1}) and (%
\ref{nonltr}).

Nonhomogeneous QC structures with mixed discrete parameters and continuous
degrees of freedom appear in a broader theoretical context related to
quantum-gravity corrections and from the point of view of an exact
renormalisation-group analysis. We omit such considerations in this work by
note that inflation in our MGTs models can be generated for $1\ll \varsigma $
and $\mathcal{M}\ll M_{P},$ which corresponds to an effective quasicrystal
potential with magnitude $\ ^{\mathbf{q}}V\ll M_{P}^{4},$ see details and
similar calculations in \cite{mavromatos}. In our approach, such values are
for nontrivial QC configurations with diagonal limits. At certain nontrivial
values $\ ^{\mathbf{q}}\varphi ,$ when $\kappa ^{-1}$ $\ ^{\mathbf{q}%
}\varphi $ are large compared to the Planck scale, a potential $\ ^{\mathbf{q%
}}V=\ ^{\mathbf{\varphi }}V(\ ^{\mathbf{q}}\mathbf{\varphi })$ (\ref%
{starobpot}) is effectively sufficiently flat to produce phenomenologically
acceptable inflation. In this model, the QC configuration determined by $\ ^{%
\mathbf{q}}\mathbf{\varphi }$ plays the role of scalar field. This
configuration determines a region with positive-definite Starobinsky
potential where the term $\exp [-\sqrt{2/3}\ ^{\mathbf{q}}\mathbf{\varphi }]$
is dominant.

In general, a nontrivial QC gravitational and effective scalar
configurations may result via generic off--diagonal parametric interactions
described by solutions type\textit{\ }(\ref{dmetrgsol}) and (\ref{lcqel}) in
effective potentials $\ ^{\mathbf{q}}V$ with constants different from (\ref%
{starobpot}), with $Q\neq \varsigma ^{2},\varpi \neq 2$ and $P\neq \sqrt{2/3}%
,$ when
\begin{equation*}
\ ^{\mathbf{q}}V=Q(1-\varpi e^{-P\ ^{\mathbf{q}}\mathbf{\varphi }}+...),
\end{equation*}%
where dots represent possible higher-order terms like $\mathcal{O}(e^{-2P\ ^{%
\mathbf{q}}\mathbf{\varphi }}).$ This means that inflation can be generated
by various types of effective quasicrystal structures which emphasizes the
generality of the model. Possible cosmological implications of QCs can be
computed following standard expressions in the slow-roll approximation for
inflationary observables (we put left labels $q$ in order to emphasize their
effective QC origin). We have%
\begin{eqnarray*}
\ ^{\mathbf{q}}\epsilon &=&\frac{M_{P}^{2}}{16\pi }\left( \frac{\ \partial \
^{\mathbf{\varphi }}V/\partial \varphi }{\ ^{\mathbf{\varphi }}V}\mid _{\ ^{%
\mathbf{q}}\mathbf{\varphi }}\right) ^{2},\ ^{\mathbf{q}}\eta =\frac{%
M_{P}^{2}}{8\pi }\frac{\ \partial ^{2}\ ^{\mathbf{\varphi }}V/\partial
^{2}\varphi }{\ ^{\mathbf{\varphi }}V}\mid _{\ ^{\mathbf{q}}\mathbf{\varphi }%
}, \\
\ ^{\mathbf{q}}n_{s} &=&1-6\ ^{\mathbf{q}}\epsilon +2\ ^{\mathbf{q}}\eta ,\
\ ^{\mathbf{q}}r=16\ ^{\mathbf{q}}\epsilon .
\end{eqnarray*}%
The a e-folding number for the inflationary phase
\begin{equation*}
\ ^{\mathbf{q}}N_{\star }=-\frac{8\pi }{M_{P}^{2}}\int\limits_{\ ^{\mathbf{q}%
}\mathbf{\varphi }_{(i)}}^{\ ^{\mathbf{q}}\mathbf{\varphi }_{(e)}}d\varphi
\frac{\ ^{\mathbf{\varphi }}V}{\partial \ ^{\mathbf{\varphi }}V/\partial
\varphi }
\end{equation*}%
with $\ ^{\mathbf{q}}\mathbf{\varphi }_{(i)}$ and $\ ^{\mathbf{q}}\mathbf{%
\varphi }_{(e)}$ being certain values of QC modifications at the beginning
and, respectively, end of inflation. At leading order, considering the small
quantity $e^{-P\ ^{\mathbf{q}}\mathbf{\varphi }},$ one computes $\ ^{\mathbf{%
q}}N_{\star }=e^{P\ ^{\mathbf{q}}\mathbf{\varphi }}/P^{2}\varpi $ yielding
\begin{equation*}
\ ^{\mathbf{q}}n_{s}=1-2P^{2}\varpi e^{-P\ ^{\mathbf{q}}\mathbf{\varphi }%
}\simeq 1-2/\ ^{\mathbf{q}}N_{\star }\mbox{ and }\ \ ^{\mathbf{q}%
}r=8P^{2}\varpi ^{2}e^{-2P\ ^{\mathbf{q}}\mathbf{\varphi }}\simeq 8/P^{2}\ ^{%
\mathbf{q}}N_{\star }^{2}.
\end{equation*}%
As a result, we get a proof that we can elaborate Starobinsky like scenarios
using generic off-diagonal gravitational configurations (in GR and/or MGTs)
determined by QC generating functions. For $\ ^{\mathbf{q}}N_{\star }=54\pm
6 $ for $P=\sqrt{2/3},$ we obtain characteristic predictions $\ ^{\mathbf{q}%
}n_{s}\simeq 0.964$ and $\ ^{\mathbf{q}}r\simeq 0.0041$ in a form highly
consistent with the Planck data \cite{plank}.

Finally, we note that for different QC configurations we may deviate from
such characteristic MGTs predictions but still remain in GR via off-diagonal
interactions resulting in QC structures. Such scenarios could not be
involved in cosmology \cite{albrecht} even the authors \cite%
{steinh1,steinh2,steinh3} made substantial contributions both to the
inflationary cosmology and physics of quasicrystals. The main problem was
that nonlinearities and parametric off-diagonal interactions were eliminated
from research from the very beginning in \cite{much,guth,linde,albrecht}
considering only the FLRW ansatz.

\subsection{Reconstructing cosmological quasicrystal structures}

We consider a model with Lagrange density (\ref{actqmg}) for $~^{q}\mathbf{f}%
(\ \widehat{\mathbf{R}})=\ \widehat{\mathbf{R}}^{2}+\mathbf{M}(~^{q}\mathbf{%
T),}$ where $~^{q}\mathbf{T}$ is the trace of the energy--momentum tensor
for an effective QC-structure determined by $(\mathbf{g}_{\alpha \beta },%
\mathbf{D}_{\mu },\varphi ).$ Let us denote $~^{q}\mathbf{M:=}\partial
\mathbf{M/\partial }~^{q}\mathbf{T}$ and $\widehat{H}:=\partial _{t}\widehat{%
a}/\widehat{a}$ for a limit $\widehat{a}(x^{i},\widehat{t})\rightarrow
\widehat{a}(t)$ in (\ref{defdm}). In general, cosmological solutions are
characterized by nonlinear symmetries (\ref{nonltr}) of generating functions
and sources when the value $\widehat{a}(t)$ is different from $\mathring{a}%
(t)$ for a standard FLRW cosmology.

To taste the cosmological scenarios one considers the redshift $1+z=\widehat{%
a}^{-1}(t)$ for a function $~^{q}T=~^{q}T(z)$ and a new \textquotedblleft
shift\textquotedblright\ derivative when $\partial _{t}s=-(1+z)H\partial
_{z},$for instance, for a function $s(t).$ Following the \ method with
nonholonomic variables elaborated in \cite{eegrg}, we obtain for QC
structures a set of three equations
\begin{eqnarray}
3\widehat{H}^{2}+\frac{1}{2}[~^{q}\mathbf{f}(z)+\mathbf{M}(z)]-\kappa
^{2}\rho (z) &=&0,  \notag \\
-3\widehat{H}^{2}+(1+z)\widehat{H}(\partial _{z}\widehat{H})-\frac{1}{2}%
\{~^{q}\mathbf{f}(z)+\mathbf{M}(z)+3(1+z)\widehat{H}^{2} &=&0,  \label{ceq1}
\\
\rho (z)\ \partial _{z}\ \mathbf{f} &=&0.  \notag
\end{eqnarray}
Using transforms of type (\ref{nonltr}) for the generating function, we fix $%
\partial _{z}\ ^{q}\mathbf{M}(z)=0$ and $\partial _{z}\ \mathbf{f}=0$ which
allow nonzero densities in certain adapted frames of references. The
functional $\mathbf{M}(~^{q}\mathbf{T)}$ encodes QC gravitational
configurations for the evolution of the energy-density of type $\rho =\rho
_{0}a^{-3(1+\vartheta )}=\rho _{0}(1+z)a^{3(1+\vartheta )}$ for the dust
matter approximation with a constant $\vartheta $ and $\rho \sim (1+z)^{3}.$

Using (\ref{ceq1}), it is possible to develop reconstruction procedures for
nontrivial QC configurations generalizing MGTs in nonholonomic variables. \
We can introduce the \textquotedblleft e-folding\textquotedblright\ variable
$\chi :=\ln a/a_{0}=-\ln (1+z)$ instead of the cosmological time $t$ and
compute $\partial _{t}s=\widehat{H}\partial _{\chi }s$ for any function $s.$
In N-adapted frames, we derive the nonholonomic field equation corresponding
to the first FLRW equation is
\begin{equation*}
~^{q}\mathbf{f}(\ \widehat{\mathbf{R}})=(\widehat{H}^{2}+\widehat{H}\
\partial _{\chi }\widehat{H})\partial _{\chi }[~^{q}\mathbf{f}(\ \widehat{%
\mathbf{R}})]-36\widehat{H}^{2}\left[ 4\widehat{H}+(\partial _{\chi }%
\widehat{H})^{2}+\widehat{H}\partial _{\chi \chi }^{2}\widehat{H}\right]
\partial _{\chi \chi }^{2}~^{q}\mathbf{f}(\ \widehat{\mathbf{R}})\mathbf{]+}%
\kappa ^{2}\rho .
\end{equation*}%
Introducing an effective quadratic Hubble rate, $\tilde{\kappa}(\chi ):=%
\widehat{H}^{2}(\chi ),$ where $\chi =\chi (\widehat{\mathbf{R}})$ for
certain parameterizations, this equation transforms into
\begin{equation}
~^{q}\mathbf{f}=-18\tilde{\kappa}(\widehat{\mathbf{R}})[\partial _{\chi \chi
}^{2}\tilde{\kappa}(\chi )+4\partial _{\chi }\tilde{\kappa}(\chi )]\frac{%
\partial ^{2}~^{q}\mathbf{f}}{\partial \widehat{\mathbf{R}}^{2}}+6\left[
\tilde{\kappa}(\chi )+\frac{1}{2}\partial _{\chi }\tilde{\kappa}(\chi )%
\right] \frac{\partial ~^{q}\mathbf{f}}{\partial \widehat{\mathbf{R}}}+2\rho
_{0}a_{0}^{-3(1+\vartheta )}a^{-3(1+\vartheta )\chi (\widehat{\mathbf{R}})}.
\label{flem}
\end{equation}%
Off-diagonal cosmological metrics encoding QC structures are of type (\ref%
{defdm}) with $t\rightarrow \chi ,$ and a functional $~^{q}\mathbf{f}(\
\widehat{\mathbf{R}})$ used for computing the generating source $\mathbf{%
\Upsilon }$ for prescribed generating function $\Phi .$ Such nonlinear
systems can be described effectively by the field equations for an
(nonholonomic) Einstein space $\widehat{\mathbf{R}}_{\ \beta }^{\alpha }=%
\widetilde{\Lambda }\delta _{\ \beta }^{\alpha }.$ The functional $\partial
~^{q}\mathbf{f/}\partial \widehat{\mathbf{R}}$ and higher functional
derivatives vanish for any functional dependence $\mathbf{f}(\widetilde{%
\Lambda })$ because $\partial _{\chi }\widetilde{\Lambda }=0.$ The
recovering procedure can be simplified substantially by using re-definitions
of generating functions.

Let us consider an example with explicit reconstruction of MGT and
nonholonomically deformed Einstein spaces with QC structure when the $%
\Lambda $CDM era can be reproduced. \textit{\ }We choose any $\widehat{a}%
(\chi )$ and $\widehat{H}(\chi )$ for an off-diagonal (\ref{defdm}). We
obtain an analog of \ the FLRW equation for ${\Lambda }$CDM cosmology,
\begin{equation*}
3\kappa ^{-2}\widehat{H}^{2}=3\kappa ^{-2}H_{0}^{2}+\rho _{0}\widehat{a}%
^{-3}=3\kappa ^{-2}H_{0}^{2}+\rho _{0}a_{0}^{-3}e^{-3\chi },
\end{equation*}%
where $H_{0}$ and $\rho _{0}$ are constant values. The effective quadratic
Hubble rate and the modified scalar curvature, $\widehat{\mathbf{R}}$, are
computed respectively,
\begin{equation*}
\tilde{\kappa}(\zeta ):=H_{0}^{2}+\kappa ^{2}\rho _{0}a_{0}^{-3}e^{-3\chi }%
\mbox{ and }\widehat{\mathbf{R}}=3\partial _{\chi }\tilde{\kappa}(\chi )+12%
\tilde{\kappa}(\chi )=12H_{0}^{2}+\kappa ^{2}\rho _{0}a_{0}^{-3}e^{-3\chi }.
\end{equation*}%
The equation (\ref{flem}) transforms into%
\begin{equation*}
X(1-X)\frac{\partial ^{2}~^{q}\mathbf{f}}{\partial X^{2}}%
+[z_{3}-(z_{1}+z_{2}+1)X]\frac{\partial ~^{q}\mathbf{f}}{\partial X}%
-z_{1}z_{2}~^{q}\mathbf{f}=0,
\end{equation*}%
for constants subjected to the conditions $z_{1}+z_{2}=z_{1}z_{2}=-1/6$ and $%
z_{3}=-1/2,$ when $3\chi =-\ln [\kappa ^{-2}\rho _{0}^{-1}a_{0}^{3}(\widehat{%
\mathbf{R}}-12H_{0}^{2})]$ and $X:=-3+\widehat{\mathbf{R}}/3H_{0}^{2}.$ The
solutions of such nonholonomic QC equations with constant coefficients and
for different types of scalar curvatures can be constructed similarly to
\cite{eegrg}. In terms of Gauss hypergeometric functions, $~^{q}\mathbf{f}%
=~^{q}F(X):=~^{q}F(z_{1},z_{2},z_{3};X),$ we obtain \
\begin{equation*}
F(X)=KF(z_{1},z_{2},z_{3};X)+BX^{1-z_{3}}F(z_{1}-z_{3}+1,z_{2}-z_{3}+1,2-z_{3};X)
\end{equation*}%
for some constants $K$ and $B.$ Such reconstructing formulas prove in
explicit form that MGT and GR theories with QC structure encode ${\Lambda }$%
CDM scenarios without the need to postulate the existence of an effective
cosmological constant. Such a constant can be stated by nonlinear
transformations and redefinitions of the generating functions and
(effective) energy momentum source for matter fields.

\section{Quasicrystal Models for Dark Energy and Dark Matter}

\label{sec5} The modern cosmological paradigm is constructed following
observational evidences that our Universe experiences an accelerating
expansion \cite{plank}. Respectively, the dark energy, DE, and dark matter,
DM, are considered to be responsible for acceleration and the dynamics of
spiral galaxies. In order to solve this puzzle of gravity, particle physics
and cosmology a number of approaches and MGTs were developed during last 20
years, see reviews and original results in \cite%
{acelun,nojod1,capoz,clifton,gorbunov,linde2,bambaodin,sami,mimet2,odints2,guend1,guend2}%
. In a series of our recent works \cite{cosmv1,cosmv2,eegrg,cosmv3}, we
proved that DE and DM effects can be modelled by generic off--diagonal
gravitational and matter field interactions both in GR and MGTs. For models
with QC structure, we do not need to "reconsider" the cosmological constant
for gravitational field equations. We suppose that an effective $\widetilde{%
\Lambda }$ can induce nonlinear symmetries of the generating functions and
effective sources resulting in QC Starobinsky like scenarios. In this
section, we prove that QC structures can be also responsible for Universe
acceleration and DE and DM effects.

\subsection{Encoding off-diagonal QC structures into canonical d--torsions}

It is possible to reformulate GR with the LC--connection in terms of an
equivalent parallel theory with the Weizenboock connection (see, for
instance, \cite{tp1,tp2,tp3}) and study $f(T)$-theories of gravity, with $T$
from torsion, which can be incorporated into a more general approach for
various modifications of the gravitational Lagrangian $R\rightarrow $ $%
f(R,T,F,L...).$ Such models can be integrated in very general forms for
geometric variables of type $(\mathbf{g,N,}\widehat{\mathbf{D}}).$
Off--diagonal configurations on GR and MGT with nonholonomic and aperiodic
structures of QC or other (noncommutative, fractional, diffusion etc.) ones
can be encoded respectively into the torsion, $\widehat{\mathcal{T}}^{\alpha
},$ and curvature, $\widehat{\mathcal{R}}_{~\beta }^{\alpha },$ tensors. By
definition, such values are defined and denoted respectively $~^{q}\mathcal{T%
}^{\alpha }:=\widehat{\mathcal{T}}^{\alpha }[~^{q}\Psi ]$ and $~^{q}\mathcal{%
R}^{\alpha }:=\widehat{\mathcal{R}}^{\alpha }[~^{q}\Psi ]$ in order to
emphasize the QC structure of generating functions and sources. Such values
can be computed in N-adapted form using the canonical d--connection 1--form $%
~^{q}\mathbf{\Gamma }_{\ \beta }^{\alpha }=\widehat{\mathbf{\Gamma }}_{\
\beta \gamma }^{\alpha }\mathbf{e}^{\gamma },$ where $~^{q}\mathbf{D}=\{~^{q}%
\widehat{\mathbf{\Gamma }}_{\ \beta \gamma }^{\alpha }\},$
\begin{eqnarray*}
~^{q}\mathcal{T}^{\alpha } &:&=~^{q}\mathbf{De}^{\alpha }=d\mathbf{e}%
^{\alpha }+~^{q}\mathbf{\Gamma }_{\ \beta }^{\alpha }\wedge \mathbf{e}%
^{\beta }=~^{q}\mathbf{T}_{\ \beta \gamma }^{\alpha }\mathbf{e}^{\beta
}\wedge \mathbf{e}^{\gamma }\mbox{\ and } \\
~^{q}\mathcal{R}_{~\beta }^{\alpha }:= &&~^{q}\mathbf{D}~^{q}\mathbf{\Gamma }%
_{\ \beta }^{\alpha }=d~^{q}\mathbf{\Gamma }_{\ \beta }^{\alpha }-~^{q}%
\mathbf{\Gamma }_{\ \beta }^{\gamma }\wedge ~^{q}\mathbf{\Gamma }_{\ \gamma
}^{\alpha }=~^{q}\mathbf{R}_{\ \beta \gamma \delta }^{\alpha }\mathbf{e}%
^{\gamma }\wedge \mathbf{e}^{\delta }.
\end{eqnarray*}%
In such formulas, we shall omit the left label "q" and write, for instance, $%
\mathbf{D,}$ $\widehat{\mathbf{\Gamma }}_{\ \beta \gamma }^{\alpha },~%
\mathbf{T}_{\ \beta \gamma }^{\alpha }$ etc. if certain continuous limits
are considered for the generating functions/sources and respective geometric
objects. Hereafter we shall work with standard N--adapted canonical values
of metrics, frames and connections which are generated by aperiodic QC $%
(~^{q}\Psi ,\ ~^{q}\Upsilon )\leftrightarrow (~^{q}\Phi ,~^{q}\widetilde{%
\Lambda }).$ Such configurations come with a nontrivial effective $~^{q}%
\mathcal{T}^{\alpha }$ induced nonholonomically. Even at the end we can
extract LC--configurations by imposing additional nonholonomic constraints
and integral sub-varieties with $(~^{q}\check{\Psi},\ ~^{q}\Upsilon
)\leftrightarrow (~^{q}\check{\Phi},~^{q}\widetilde{\Lambda })$ all diagonal
and off-diagonal cosmological solutions are determined by geometric and
physical data encoded in $\{~^{q}\mathbf{T}_{\ \beta \gamma }^{\alpha }\}.$
For $~^{q}\mathbf{D}\rightarrow ~^{q}\nabla ,$ the gravitational and matter
field interactions are encoded into $\mathbf{e}_{\ \alpha ^{\prime
}}^{\alpha }[\mathbf{g,N}]$ like in (\ref{odans}). We can work
in equivalent form with different type theories when
 $R\Longleftrightarrow \widehat{\mathbf{R}}\Longleftrightarrow f(~^{q}\mathbf{R%
})\Longleftrightarrow f(~^{q}\mathcal{T})$ 
are all completely defined by the same metric structure and data $(\mathbf{%
g,N).}$ Here we note that $~^{q}\mathbf{T}$ is constructed for the canonical
d--connection $~^{q}\mathbf{D}$ in a metric--affine spacetime with aperiodic
order and this should be not confused with theories of type $f(R,T),$ where $%
T$ is for the trace of the energy-momentum tensor.

We construct an equivalent $f(~^{q}\mathcal{T})$ theory for DE and DM
configurations determined by a QC structure: Let us consider
respectively the contorsion and quasi-contorsion tensors
\begin{equation*}
~^{q}\mathbf{K}_{\ \quad \lambda }^{\mu \nu }=\frac{1}{2}\left( ~^{q}\mathbf{%
T}_{\ \quad \lambda }^{\mu \nu }-~^{q}\mathbf{T}_{\ \quad \lambda }^{\nu \mu
}+~^{q}\mathbf{T}_{\lambda \quad }^{\quad \nu \mu }\right) ,\ ^{q}\mathbf{S}%
_{\lambda \quad }^{\quad \nu \mu }=\frac{1}{2}\left( ~^{q}\mathbf{K}_{\
\quad \lambda }^{\mu \nu }+\delta _{\lambda }^{\mu }~^{q}\mathbf{T}_{\ \quad
\alpha }^{\alpha \nu }-\delta _{\lambda }^{\nu }~^{q}\mathbf{T}_{\ \quad
\alpha }^{\alpha \mu }\right)
\end{equation*}%
for any $~^{q}\mathcal{T}^{\alpha }=\{~^{q}\mathbf{T}_{\ \quad \lambda
}^{\mu \nu }\}.$ Then the canonical torsion scalar is defined $~^{q}\mathcal{%
T}:=~^{q}\mathbf{T}_{\ \beta \gamma }^{\alpha }~^{q}\mathbf{T}_{\alpha \quad
}^{\quad \beta \gamma }.$ The nonholonomic redefinition of actions and
Lagrangians (\ref{actqmg}) and (\ref{mgract}) in terms of $~^{q}\mathcal{T}$
$\ $is
\begin{equation}
\mathcal{S}=\int d^{4}u\left[ \frac{\ _{G}^{q}\mathcal{L}}{2\kappa ^{2}}+\
^{m}\widehat{\mathcal{L}}\right]  \label{tpact}
\end{equation}%
where the Lagrange density for QC gravitational interactions is $\ _{G}^{q}%
\mathcal{L}=~^{q}\mathcal{T}+f(~^{q}\mathcal{T}).$

The equations of motion in a flat FLRW universe derived for solutions of
type (\ref{defdm}) (for simplicity, we omit small parameter off-diagonal
deformations (\ref{smalld})) are written in the form
\begin{eqnarray*}
6H^{2}+12H^{2}f^{\circ }(~^{q}\mathcal{T})+f(~^{q}\mathcal{T}) &=&2\kappa
^{2}\rho _{\lbrack i]}, \\
2(2\partial _{t}H+3H^{2})+f(~^{q}\mathcal{T})+4(\partial
_{t}H+3H^{2})f^{\circ }(~^{q}\mathcal{T})-48H^{2}(\partial _{t}H)f^{\circ
\circ }(~^{q}\mathcal{T}) &=&2\kappa ^{2}p_{[i]},
\end{eqnarray*}%
where $f^{\circ }:=df/d$ $~^{q}\mathcal{T}$ and $\rho _{\lbrack i]}$ and $%
p_{[i]}$ respectively denote the energy and pressure of a perfect fluid
matter embedded into a QC like gravitational and scalar matter type
structure. Similar equations have been studied in Refs. \cite%
{df1,df2,df3,df4}. For $\kappa ^{2}=8\pi G,$ these equations can be written
respectively as constraints equations
\begin{equation*}
3H^{2}=\rho _{\lbrack i]}+~^{q}\rho ,2\partial _{t}H=-(\rho _{\lbrack
i]}+p_{[i]}+~^{q}\rho +~^{q}p),
\end{equation*}%
with additional effective QC type matter%
\begin{equation}
~^{q}\rho =-6H^{2}f^{\circ }-f/2,\ ~^{q}p=2\partial _{t}H(1+f^{\circ
}-12H^{2}f^{\circ \circ })+6H^{2}f^{\circ }+f/2.  \label{qcenpres}
\end{equation}%
Now we can further develop our approach with DE and DM determined by
aperiodic QC configurations of gravitational - scalar field systems.%
\footnote{%
Above equations can be written in a standard form for $f$--modified
cosmology with
 $\ ^{ef}\Omega =\Omega _{\lbrack i]}+~^{q}\Omega :=\frac{\rho _{\lbrack i]}}{%
3H^{2}}+\frac{~^{q}\rho }{3H^{2}}=1$, 
 for  effective $~^{ef}\rho =\rho _{\lbrack i]}+~^{q}\rho
,~^{ef}p=p_{[i]}+~^{q}p$ and $~^{ef}\omega =~^{ef}p/~^{ef}\rho $ $\ $%
encoding an aperiodic QC order.}

\subsection{Interaction between DE and DM in aperiodic QC vacuum}

In this section, we ignore all other forms of energy and matter and study
how DE and DM interact in a directly aperiodically QC structured. Respective
densities of QC dark energy and dark matter are parameterized
 $\ ^{q}\rho =~_{DE}^{q}\rho +~_{DM}^{q}\rho $  and $\ ^{q}p=~_{DE}^{q}p+~_{DM}^{q}p$, 
 when (\ref{qcenpres}) is written in the form
\begin{equation*}
2\partial _{t}(~_{DE}^{q}\rho +~_{DM}^{q}\rho )=(\partial _{t}~^{q}\mathcal{%
T)}\left( f^{\circ }+2~^{q}\mathcal{T}f^{\circ \circ }\right) .
\end{equation*}%
For perfect two fluid models elaborated in N--adapted form \cite%
{cosmv1,cosmv2,eegrg,cosmv3}, the interaction DE and DM equations are written%
\begin{eqnarray}
\kappa ^{2}(~^{q}\rho +~^{q}p) &=&-2\partial _{t}H,\mbox{ subjected to  }
\label{constreq} \\
\partial _{t}(~_{DE}^{q}\rho )+3H(~_{DE}^{q}p+~_{DE}^{q}\rho ) &=&-Q%
\mbox{
and }\partial _{t}(~_{DM}^{q}\rho )+3H(~_{DM}^{q}p+~_{DM}^{q}\rho )=Q.
\notag
\end{eqnarray}%

Above equations result in the following functional equation
\begin{equation*}
2~^{q}\mathcal{T}f^{\circ \circ }+f^{\circ }+1=0,
\end{equation*}%
which can be integrated in trivial and nontrivial forms with certain
integration constants $C,C_{0}$ and $C_{1}=0$ (this condition follows from (%
\ref{constreq})),%
\begin{equation*}
f(~^{q}\mathcal{T})=\left\{
\begin{array}{c}
-~^{q}\mathcal{T}+C \\
-~^{q}\mathcal{T}-2C_{0}\sqrt{-~^{q}\mathcal{T}}+C_{1}%
\end{array}%
\right. .
\end{equation*}%
So, the QC structure effectively contributes to DE and DM interaction via a
nontrivial nonholonomically induced torsion structure. Such nontrivial
aperiodic configurations exist via nontrivial $C$ and $C_{0}$ even we impose
the conditions $~^{q}\mathcal{T}$ in order to extract certain diagonal
LC--configurations. We note that in both \ cases of solutions for $f(~^{q}%
\mathcal{T})$ we preserve the conditions $~^{ef}\Omega =1$ and $~^{ef}\omega
=-1.$

\subsection{Quasicrystal DE structures and matter sources}

We analyse how aperiodic QC structure modify DE and DM and ordinary matter
OM interactions and cosmological scenarios, see similar computations in \cite%
{df3,df4} but for a different type of torsion (for the Weitzenb\"{o}ck
connection).

\subsubsection{Interaction between DE and ordinary matter in gravitational
QC media}

Now, we model a configuration when aperiodic DE interacts with OM (we use
label "o" from ordinary, $(~^{o}\rho +~^{o}p)$ for $~^{q}\rho
=~_{DE}^{q}\rho .$ We obtain such equations of interactions between DE and
DM equations are written%
\begin{eqnarray}
\kappa ^{2}(~_{DE}^{q}\rho +~_{DE}^{q}p+~^{o}\rho +~^{o}p) &=&-2\partial
_{t}H,\mbox{ subjected to  }  \label{qcdeom} \\
\partial _{t}(~_{DE}^{q}\rho )+3H(~_{DE}^{q}p+~_{DE}^{q}\rho ) &=&-Q%
\mbox{
and }\partial _{t}(~^{o}\rho )+3H(~^{o}\rho +~^{o}p)=Q.  \notag
\end{eqnarray}%
The equation (\ref{qcenpres}) transform into
\begin{equation*}
\partial _{t}(~_{DE}^{q}\rho )=(\partial _{t}~^{q}\mathcal{T})(~^{q}\mathcal{%
T}f^{\circ \circ }+\frac{1}{2}f^{\circ }),
\end{equation*}%
which together with above formulas result in
\begin{equation*}
\partial _{t}(~_{DE}^{q}\rho +~^{o}\rho +\frac{1}{2}~^{q}\mathcal{T})=0.
\end{equation*}%
For $f(~^{q}\mathcal{T}),$ these formulas result in a second order
functional equation
\begin{equation*}
(2~^{q}\mathcal{T}f^{\circ \circ }+f^{\circ }+1)=-2(~^{o}\rho )^{\circ }.
\end{equation*}%
We can construct solutions of this equation by a splitting into two
effective ODEs with a nonzero constant $Z_{0},$ when
\begin{equation*}
f^{\circ \circ }+(2~^{q}\mathcal{T})^{-1}f^{\circ }=-Z_{0}\mbox{ and }%
2~^{o}\rho +1=2Z_{0}~^{q}\mathcal{T}.
\end{equation*}%
Such classes of solutions are determined by integration constants $C_{2}$
and $C_{3}=2C_{4}$ (this condition is necessary in order to solve (\ref%
{constreq})): for $~^{o}\rho =-C_{4}-~^{q}\mathcal{T}+Z_{0}(~^{q}\mathcal{T}%
)^{2}/2,$ the aperiodic QC contribution is
\begin{equation*}
f(~^{q}\mathcal{T})=C_{3}-2C_{2}\sqrt{|-~^{q}\mathcal{T}|}-Z_{0}(~^{q}%
\mathcal{T})^{2}/3.
\end{equation*}%
We can choose $H_{0}=74.2\pm 3.6\frac{Km}{s}\frac{Mp}{c}$ and $t_{0}$ as the
present respective Hubble parameter and cosmic time and state the current
density of the dust $\rho (t_{0})=~^{m}\rho _{0}=3\times 1.5\times
10^{-67}eV^{2}.$ For an arbitrary constant $C_{2},$ we get the gravitational
action (\ref{constreq}) and $~^{o}\rho $ both modified by QC contributions
via
\begin{equation*}
C_{4}=~^{m}\rho _{0}-3H_{0}^{2}(1-6Z_{0}H_{0}^{2}).
\end{equation*}%
The effective parameters of state
\begin{equation*}
~^{ef}\omega =-(~^{q}\mathcal{T})^{-1}\{Z_{0}(~^{q}\mathcal{T}%
)^{2}+4[1-2\partial _{t}H~Z_{0}(~^{q}\mathcal{T})]+C_{3}\}\mbox{ and }%
~^{ef}\Omega =1
\end{equation*}%
describe an universe dominated by QC dark energy interacting with ordinary
matter.

\subsubsection{Van der Waals fluid interacting with aperiodic DM}

The state equation for such a fluid (with physical values labeled by $w$) is
\begin{equation*}
~^{w}p(3-~^{w}\rho )+8~^{w}p~^{w}\rho -3(~^{w}\rho )^{2}=0,
\end{equation*}%
which results in the equations for interaction of the QC DE with such a van
der Waals OM,
\begin{eqnarray*}
\kappa ^{2}(~_{DE}^{q}\rho +~_{DE}^{q}p+~^{w}\rho +~^{w}p) &=&-2\partial
_{t}H,\mbox{ subjected to  } \\
\partial _{t}(~_{DE}^{q}\rho )+3H(~_{DE}^{q}p+~_{DE}^{q}\rho ) &=&-Q%
\mbox{
and }\partial _{t}(~^{w}\rho )+3H(~^{w}\rho +~^{w}p)=Q.
\end{eqnarray*}%
Such equations are similar to (\ref{qcdeom}) but with the OM pressure and
density subjected to another state equation and modified DE interaction
equations. The solutions the aperiodic QC contribution can be constructed
following the same procedure with two ODEs and expressed for $~^{w}\rho
=-C_{5}+Z_{0}(~^{q}\mathcal{T})^{2}/2,C_{7}=2C_{5},$ as
\begin{equation*}
f(~^{q}\mathcal{T})=C_{7}+C_{6}\sqrt{|-~^{q}\mathcal{T}|}/2-~^{q}\mathcal{T}%
-Z_{0}(~^{q}\mathcal{T})^{2}/3.
\end{equation*}%
Taking $\partial _{t}H(t_{0})=0$ and $~_{DE}^{q}p(t_{0})+$ $~_{DE}^{q}\rho
(t_{0})=0,$ which constrains (see above equations) $~^{w}p(t_{0})+~^{w}\rho
(t_{0})=0,$ and results in
\begin{equation*}
C_{5}=3Z_{0}H_{0}^{2}+|74-96~^{w}\omega |^{1/2}+\frac{5}{3},
\end{equation*}%
for typical values $~^{w}\omega =0.5$ and $E=10^{-10}.$

\subsubsection{Chaplygin gas and DE - QC configurations}

Another important example of OM studied in modern cosmology (see, for
instance, \cite{kamen}) is that of Chaplygin, $ch$, gas characterized by an
equation of state $~^{ch}p=-Z_{1}/~^{ch}\rho ,$ for a constant $Z_{1}>0.$
The corresponding equations for interactions between DE and such an OM is
given by
\begin{eqnarray*}
\kappa ^{2}(~_{DE}^{q}\rho +~_{DE}^{q}p+~^{ch}\rho +~^{ch}p) &=&-2\partial
_{t}H,\mbox{ subjected to  } \\
\partial _{t}(~_{DE}^{q}\rho )+3H(~_{DE}^{q}p+~_{DE}^{q}\rho ) &=&-Q%
\mbox{
and }\partial _{t}(~^{ch}\rho )+3H(~^{ch}\rho +~^{ch}p)=Q.
\end{eqnarray*}%
The solutions for this system can be written for $~^{ch}\rho
=C_{8}+Z_{0}(~^{q}\mathcal{T})^{2}/2,C_{10}=2C_{8},$ as
\begin{equation*}
f(~^{q}\mathcal{T})=C_{10}+C_{9}\sqrt{|-~^{q}\mathcal{T}|}/2-~^{q}\mathcal{T}%
-Z_{0}(~^{q}\mathcal{T})^{2}/3.
\end{equation*}%
Let us assume $\partial _{t}H(t_{0})=0$ and $~_{DE}^{q}p(t_{0})+$ $%
~_{DE}^{q}\rho (t_{0})=0,$ which results in $~^{ch}p(t_{0})+~^{ch}\rho
(t_{0})=0$ and
\begin{equation*}
C_{8}=18Z_{0}H_{0}^{4}+|Z_{1}-9Z_{0}H_{0}^{4}(1+36Z_{0}H_{0}^{4})|^{1/2},
\end{equation*}%
for typical values $Z_{1}=1$ and $E=10^{-10}.$

The solutions for different types of interactions of QC like DE and DM with
OM subjected to corresponding equations of state (for instance, of van der
Waals or Chaplygin gase) prove that aperiodic spacetime strucutres result,
in general, in off--diagonal cosmological scenarios which, in diagonal
limits, result in effects related directly to terms containing contributions
of nonholonomically induced torsion. If the constructions are redefined in
coordinate type variables, such terms transform into certain generic
off--diagonal coefficients of metrics.

\section{Discussion and Conclusions}

\label{sec6}This paper is devoted to the study of aperiodic quasicrystal
like gravitational and scalar field structures in acceleration cosmology. It
applies certain geometric methods for constructing exact solutions in
mathematical cosmology. The main conclusion is that exact solutions with
aperiodic order in modified gravity theories, MGTs, and general relativity,
GR, and with generic off--diagonal metrics, confirm but also offer
interesting alternatives to the original Starobinsky model. Here we
emphasize that our work concerns possible spacetime aperiodic order and QC
like discrete and continuous configurations at cosmological scales. This is
different from the vast majority of QCs (discovered in 1982 \cite{qsnp},
which attracted the Nobel prize for chemistry in 2011) that are made from
metal alloys. There are also examples of QCs found in nanoparticles and
soft-matter systems with various examples of block copolymers etc. \cite%
{qcalloys1,qcalloys2,qcalloys3,qcalloys4,qcalloys5}. In a complimentary way,
it is of special interest to study configurations with aperiodic order also
present in astronomy and cosmology, when a number of observational data
confirm various types of filament and deformed QC structures, see \cite%
{crystalinks}.

In this work, the emergence of an aperiodic ordered structure in
acceleration cosmology is investigated following geometric methods of
constructing exact and parametric solutions in modified gravity theories,
MGTs, and in general relativity, GR (such methods with applications in
modern cosmology are presented in \cite{gvvafdm,cosmv1,cosmv2,eegrg}). For
instance, QC configurations may be determined by generating functions
encoding, for instance, a \ "golden rotation" of $\arccos (\tau ^{2}/2\sqrt{2%
})\approx 22.2388^{\circ }$ (where the golden ratio is given by $\tau =\frac{%
1}{2}(1+\sqrt{5})$), see \cite{klee1,klee2}. Implementation of the QSN model
of \cite{klee2}, instead of involving 30 wavevectors at two scales listed in
table 1 of \cite{subr}, is involving 108 wavevectors at two scales,
described by Dirichlet integers, solutions of the 3D/6D Dirichlet
split-sphere equation of radius $\sqrt{8}$, see \cite{bubuianu17}.

The reason to use such aperiodic and discrete parametrized generating
functions and effective sources of matter is that various cosmological
scales can be reproduced as certain nonholonomic deformations and diffusion
processes from a chosen QC configuration. The priority of our geometric
methods is that we can work both with continuous and discrete type
generating functions which allows us to study various non-trivial deformed
networks with various bounds and lengths re-arranging and deforming, for
instance, icosahedral arrangements of tetrahedra etc.

The aperiodic QC gravity framework proves to be very useful, since many
geometric and cosmological evolution scenarios can be realized in the
context of this approach. The question is whether such models can be
considered as viable ones in order to alternatively explain Starobinsky-like
scenarios and provide a physical ground for explicit models of dark energy,
DE, and dark matter,\ DM. Working with arbitrary generating functions and
effective source it seems that with such MGTs everything can be realized and
certain lack of predictability is characteristic. From our point of view,
the anholonomic frame deformation method,\ AFDM, is more than a simple
geometric methods for constructing exact solutions for certain classes of
important nonlinear systems of partial differential equations, PDEs, in
mathematical relativity. It reflects new and formerly unknown properties and
nonlinear symmetries of the (modified)\ Einstein equations when generic
off--diagonal interactions and mixed continuous and discrete structures are
considered because in vacuum and non-vacuum gravitational configurations.
The AFDM is appealing in some sense it is "economical and very efficient";
allowing us to proceed in the same manner but with fractional / random /
noncommutative sources and their respective interaction parameters. We can
speculate on the existence of noncommutative and/or nonassociative QC
generalized structures in the framework of classical MGTs. Moreover, certain
compatibility with the cosmological observational data can be achieved and,
in addition, there are developed realistic models of geometric flows and
growth of QCs related to accelerating cosmology.

An interesting and novel research stream is related to the possibility of
encoding aperiodic QC structures into certain nonholonomic and generic
off--diagonal metric configurations with nonholonomically induced canonical
torsion fields. Such alternatives to the teleparallel and other MGTs
equivalents of GR allow us to elaborate in the most "economic" way on \ QC
models for DE and DM, and study "aperiodic dark" interactions with ordinary
matter (like van der Vaals and Chaplygin gas). A procedure which allows us
to reconstruct QCs is outlined following our former nonholonomic
generalizations \cite{cosmv1,cosmv2,eegrg}. However, QC structures are
characterised by Lypunov type functionals for free energy, which for
geometric models of gravity are related to certain generalized Perelman's
functionals studied in \cite{vanp}. Following such an approach, a new theory
with DE and DM originating from aperiodic geometry and with generalized
Ricci flows with (non) holonomic/ commutative / fractional structures has to
be developed and we defer such issues to our future works \cite%
{partner2,partner3}.

\vskip5pt \textbf{Acknowledgments:}\ SV research on geometric methods for
constructing exact solutions in gravity theories and applications in modern
cosmology and astrophysics was supported by the program IDEI:
PN-II-ID-PCE-2011-3-0256, and former senior research fellowships from CERN,
and DAAD. Some results were presented at GR21 in NY. Some examples of
quasi-periodic and aperiodic cosmological solutions were elaborated
following a Consulting Agreement (for a visiting researcher) with a private
organization QGR-Topanga, California, the USA, during May 18, 2016 - June 2,
2017.

\end{document}